\documentclass[12pt]{article}
\usepackage{mathrsfs}
\usepackage{amsmath}
\usepackage{mathrsfs}
\usepackage{amssymb}
\usepackage{color}
\usepackage{psfrag}

\newcommand{\bea}{\begin{eqnarray}}
\newcommand{\eea}{\end{eqnarray}}
\newcommand{\be}{\begin{equation}}
\newcommand{\ee}{\end{equation}}
\newcommand{\vs}[1]{\vspace{#1 mm}}

\newcommand{\dsl}{\pa \kern-0.5em /}

\newcommand{\pa}{\partial}

\newcommand{\nn}{\nonumber\\}

\begin{document}
\topmargin 0mm
\oddsidemargin 0mm

\begin{flushright}

USTC-ICTS-18-02\\

\end{flushright}

\vspace{2mm}

\begin{center}

{\Large \bf Some aspects of interaction amplitudes of D branes carrying worldvolume fluxes}

\vs{10}

{\large J. X. Lu}

\vspace{4mm}

{\em
Interdisciplinary Center for Theoretical Study\\
 University of Science and Technology of China, Hefei, Anhui
 230026, China\\
 
}

\end{center}

\vs{10}

\begin{abstract}
  We report a systematic study of the stringy interaction between two sets of Dp branes placed parallel at a separation in the presence of two worldvolume fluxes for each set.   We focus in this paper on that the two fluxes on one set have the same structure as those on the other set but they in general differ in values, which can be both electric or both magnetic or one electric and one magnetic. We compute the respective stringy interaction amplitude  and find that the presence of electric fluxes gives rise to the open string pair production while that of magnetic ones to the open string tachyon mode. The interplay of these two leads to the open string pair production enhancement in certain cases when one flux is electric and the other is magnetic.  In particular,  we find that this enhancement occurs even when the electric flux and the magnetic one share one common field strength index which is impossible in the one-flux case studied previously by the present author and his collaborator in \cite{Lu:2009au}.  This type of enhancement may  have realistic physical applications, say,  as a means to explore the existence of extra dimensions.       
\end{abstract}

\newpage
\section{Introduction}
D brances are one type of non-perturbative stable Bogomol'ny-Prasad-Sommerfield  (BPS) solitonic extended objects in superstring theories (for example, see \cite{Duff:1994an}), preserving one half of the spacetime supersymmetries. These object are important or useful mainly because though they are non-perturbative, their dynamics can still be described, when the string coupling is small, by the perturbative open string with its two ends satisfying the usual Neumann Boundary conditions along the brane directions and the so-called Dirichlet boundary conditions along directions transverse to the branes \cite{Polchinski:1995mt}. When two such Dp-branes\footnote{\label{fn1}For having a distance between the two, we need to have $p \le 8$.} are placed parallel to each other at a separation and are at rest,  there is no net interaction acting between the two and this system, just like either of the Dp branes, is also a stable BPS one, preserving one half of the spacetime supersymmetries. 

A Dp brane has its tension and carries also the so-called RR charge. One therefore expects in general an attractive force due to their tensions and a repulsive one due to their RR charges between two such Dp branes. The BPS nature of the Dp brane relates the charge and the tension and as such the sum of these two contributions gives a vanishing net interaction. We can check this by computing the lowest order stringy interaction amplitude from either a closed string tree-level cylinder diagram or equivalently an open string one-loop annulus one.  In either computation, we have two contributions.  The so-called NS-NS contribution, due to the brane tension, is as expected attractive, while the so-called R-R contribution, due to the RR charges, is repulsive.  The sum of the two gives an expected zero net interaction by making use of the  usual `abstruse identity' \cite{Polchinski:1995mt}.

When each Dp-brane carries fluxes, which can be electric and/or magnetic ones\footnote{\label{fn2}The electric flux on a Dp-brane stands for the presence of F-strings, forming the so-called (F, Dp) non-threshold bound state\cite{Witten:1995im, Schmidhuber:1996fy, Arfaei:1997hb,
Lu:1999qia, Lu:1999uca, Lu:1999uv, Hashimoto:1997vz,
DiVecchia:1999uf}, while a magnetic flux stands for that of co-dimension 2 D-branes inside the original Dp brane, forming the so-called (D(p-2), Dp) non-threshold bound state\cite{Breckenridge:1996tt, Costa:1996zd, Di Vecchia:1997pr}, from the spacetime perspective. These fluxes are in general quantized.  We will not discuss their quantizations in the text for simplicity due to their irrelevance for the purpose of this paper.}, the interaction is in general non-vanishing. For large brane separation, this interaction, if non-vanishing, has to be attractive since it is due to different branes and the only contributions are from their tensions (different brane charges do not interact).  For small brane separation, the story is a bit complicated as we will see. The best description is in terms of the open string one. If there is a non-vanishing interaction,  the underlying system breaks all supersymmetries and we  expect to have some interesting physics process to occur, especially when the brane separation is small.

From the open string perspective,  the open string one-loop annulus diagram can be viewed either as a virtual open string circulating a closed time loop or as a pair of virtual open string and virtual anti open string creating from the vacuum at certain moment, existing for a short period of time and finally annihilating to the vacuum. The two ends of the virtual open string pair connecting the two Dp branes appear just as virtual charge and anti-charge pair to each Dp brane. If the added fluxes on each Dp contain an electric one,  this electric flux can provide a force acting on the virtual charge and anti-charge pair to pull them apart, which can also be understood as providing the energy needed to the virtual pair, to make them become real, i.e., the analog of the Schwinger pair production. So we expect the interaction amplitude, in the presence of electric flux(es), not only to be non-vanishing but also to have an imaginary part,  giving rise to the open string pair production.  In general, the pair production rate is vanishingly small and suppressed exponentially by the brane separation.  However, when magnetic fluxes are also present in a certain way\footnote{For D = 26 bosonic string case, magnetic fluxes were also considered in a general setting in the spirit of Bachas and Porrati \cite{Bachas:1992bh,Porrati:1993qd} in \cite{Acatrinei:2000qm}. It is not clear if there is a similar pair production enhancement uncovered here in the bosonic context (It does not appear so in the specific example given there.).},  this open string pair production rate is greatly enhanced and becomes significant to have potential physical applications.  

 We would like to stress that the present pair production in Type II superstring case is different from that in the Type I superstring case as given in \cite{Bachas:1992bh,Porrati:1993qd} .  For a single Type II Dp brane, we have  a $U(1)$ gauge group and computations give a vanishing open string annulus amplitude as well as a vanishing open string pair production rate even if the brane carries a constant worldvolume electric flux. These vanishing results are due to that the open string is oriented and is therefore charge-neutral in the sense that its two ends carry the respective U(1) charge $+1$ and $-1$ with zero-net charge.  This is also consistent with the fact that a Type II Dp brane carrying a constant electric flux is a 1/2 BPS non-threshold bound state (F, Dp) as discussed in footnote \ref{fn2}.  So this system is stable rather than unstable and the pair production cannot occur.  In order to have the open string pair production in Type II, the simplest possible choice is to consider two Dp branes placed parallel at a separation with each carrying a different electric flux. This is the rational for considering such a system of two Dp branes in this paper and as mentioned above already the open strings produced are directly related to the dimensions transverse to the branes. So a detection of the pair production by an observer living on the brane will signal the existence of extra dimensions, for exmple, for $p = 3$ case. 
 
 Given the above rational for the open string pair production in Type II string theories discussed in this paper,  the open string pair production discussed in \cite{Bachas:1992bh,Porrati:1993qd} is for the charged unoriented open string in Type I superstring with its two ends carrying their respective charge $e_{1}$ and $e_{2}$ coupled to a constant spacetime background electric field.  This background field picks up some $U(1)$ direction inside the non-abelian $SO(32)$ in Type I. 
 Since there are different choices of this $U(1)$ embedding inside $SO(32)$, this may give different values of the charge $e_{1}$ and $e_{2}$.  So for this case $e_{1} + e_{2}$ can be nonzero and as such, as discussed in  \cite{Bachas:1992bh,Porrati:1993qd}, there can be non-zero open string pair production rate.   In terms of modern D-brane language, we know that the gauge group $SO(32)$ is from the 16  D9 (spacetime filling) branes  in type I theory and the open string considered describes certain overall dynamics, characterized by the $U(1)$, of these 16 D9 branes.  Consistent with this, the two ends of Type I unoriented open string, apart from the background field directions, obey the usual Neumann boundary conditions. There is no sense of brane separation here.  The direct relevance to this case is a D9 in Type IIB.  As stressed above,  the corresponding Type IIB open string pair production rate always vanishes even if a constant worldvolume electric flux is applied.  So the physics is  different in these two cases even though there are similarities technically.   
 
 For other Type II Dp-branes with $p \le 8$,  the two ends of open string now obey Dirichlet boundary conditions along directions transverse to the brane in addition to the Neumann 
 boundary conditions along the brane directions. For a system of two Dp branes placed parallel at a separation with each carrying fluxes considered in this paper, the  exponential brane separation suppression factor, appearing in the interaction amplitude and the corresponding possible open string pair production rate, comes from the zero-mode contribution from the Dirichlet directions.    
 
 The operator structure of the boundary state for a Dp brane holds true even with the presence of general external
fluxes on the worldvolume \cite{DiVecchia:1999uf}  and using this closed string boundary state approach we can compute the closed string cylinder amplitude between the two Dp branes considered for any constant worldvolume flux, especially when certain technical tricks can be used, as discussed in the following section, to simplify the computations. The corresponding open string annulus amplitude can be obtained, for the purpose of obtaining the possible open string  pair production rate, simply by using a Jacobi transformation. This gives an advantage over the open string approach adopted in  \cite{Bachas:1992bh,Porrati:1993qd} for which only pure electric or magnetic field was considered in Type I superstring case. It appears complicated and difficult there if both electric and magnetic fields are present.    
   
 Without further ado, in this paper we will compute the interaction amplitude for a system of two sets of Dp branes, placed parallel at a separation, with each set carrying two fluxes with the same structure but different in values in the sense specified later on. We will also compute the corresponding open string pair production rate if any and discuss the relevant analytic structures of the amplitude. We will give a complete account of the aforementioned two-flux cases in this paper.  Depending on the structure of the two fluxes on each set of the Dp branes, we have three cases to consider:  1) $8 \ge p \ge 2$, 2) $8 \ge p \ge 3$ and 3) $8\ge p \ge 4$.  We will explore the nature (attractive or repulsive) of the interaction at large brane separation and small brane separation, respectively, and study various instabilities
such as the onset of tachyonic one at small brane separation. We will  determine at which conditions there exists open string pair production and its possible enhancement. We will also speculate possible applications of the enhanced open string pair production.  
   
This paper is organized as follows.  In section 2,  we provide the basic setup for the computations of various interaction amplitudes for systems considered. In section 3, we compute the interaction amplitude and give a complete  analytical analysis of this amplitude for the system of two sets of Dp branes placed parallel at a separation when the two fluxes on each set share one common field strength index. This case requires $8 \ge p \ge 2$.  Here we find a new possibility, when one flux is electric and the other is magnetic, that gives rise to a new pair production enhancement.  This possibility will not occur when each set of Dp branes carry only one flux as studied previously in \cite{Lu:2009au}. In section 4, we repeat the amplitude computation and its analysis for the system of two sets of Dp branes in a similar fashion but with the two fluxes on each set sharing no common field strength index when one is electric and the other magnetic or sharing one common field strength index when both are magnetic. This case corresponds to $8 \ge p \ge 3$.  This is the most interesting case for having the great enhancement of open string pair production when there are one electric flux and one magnetic one present on each set of the Dp branes. In particular, when the two electric fluxes are almost identical and the two magnetic fluxes are opposite in direction,  the open string pair production rate has a great enhancement which is quite unexpected and this rate is the largest when $p = 3$. As we will discuss in section 6, this is the case that has potentially realistic applications, for example, one can use this as a means to explore the existence of extra dimension(s) among other things.   In section 5, we repeat the same process for the system of two sets of Dp branes again in a similar fashion but now with the two magnetic fluxes on each set sharing no common field strength index. This corresponds to $8 \ge p \ge 4$ case.   We will discuss and conclude this paper in section 6.      

\section{The basic setup} 

In this section, we will provide the basis for computing the lowest order stringy interaction amplitude for a system of two sets of Dp branes placed parallel at a separation with each set carrying certain fluxes. For this, we consider first the closed string cylinder diagram with Dp branes represented by their respective boundary state $|B\rangle$\cite{Callan:1986bc,Polchinski:1987tu}.  For such a description, there are two sectors, namely NS-NS and R-R sectors. In each sector, we have two implementations for the boundary conditions of a Dp brane, giving two boundary states $|B, \eta \rangle$, with $\eta = \pm$. However, only the combinations 
\bea\label{gsopbs}
 &&|B \rangle_{\rm NS} = \frac{1}{2} \left[|B, + \rangle_{\rm NS} - |B, - \rangle_{\rm NS}\right],\nn
 &&|B \rangle_{\rm R} = \frac{1}{2} \left[|B, + \rangle_{\rm R} + |B, - \rangle_{\rm R}\right],
 \eea
  are selected by the  Gliozzi-Scherk-Olive (GSO) projection in the NS-NS and R-R sectors, respectively.   The boundary state $|B, \eta\rangle$ for a Dp-brane can be expressed as the product of a matter part and a ghost part \cite{Billo:1998vr, Di Vecchia:1999fx}, i.e. 
 \be\label{bs}
  |B, \eta\rangle = \frac{c_p}{2} |B_{\rm mat}, \eta\rangle |B_{\rm g}, \eta\rangle,
 \ee
 where 
 \be\label{mgbs}
 |B_{\rm mat}, \eta\rangle = |B_X \rangle|B_\psi, \eta\rangle,  \quad |B_{\rm g},\eta\rangle = |B_{\rm gh}\rangle |B_{\rm sgh}, \eta\rangle
 \ee
  and the overall normalization
$c_p = \sqrt{\pi}\left(2\pi \sqrt{\alpha'}\right)^{3 - p}. $  

As discussed in \cite{DiVecchia:1999uf}, the operator structure of the
boundary state holds true even with the presence of  external
fluxes on the worldvolume and is always of the form 
\be\label{xbs}
 |B_X\rangle ={\rm exp} (-\sum_{n =1}^\infty \frac{1}{n} \alpha_{- n} \cdot M
\cdot {\tilde \alpha}_{ - n}) |B_X\rangle_0,
\ee
 and 
 \be\label{bspsinsns}
 |B_\psi, \eta\rangle_{\rm NS} = - {\rm i}~ {\rm exp} (i \eta
\sum_{m = 1/2}^\infty \psi_{- m} \cdot M \cdot {\tilde \psi}_{- m})|0\rangle,
\ee for the NS-NS sector and 
\be\label{bspsirr}
 |B_\psi,\eta\rangle_{\rm R} = - {\rm exp} (i \eta \sum_{m = 1}^\infty
\psi_{- m} \cdot M \cdot {\tilde \psi}_{- m}) |B,
\eta\rangle_{0\rm R},
\ee for the R-R sector. The ghost boundary states are the standard ones as given in \cite{Billo:1998vr}, independent of the fluxes, which we will not present here.   The M-matrix\footnote{We have changed the previously often used symbol $S$ to the current $M$ to avoid a possible confusion with the S-matrix in scattering amplitude.}, 
the zero-modes  $|B_X\rangle_0$ and $|B,
\eta\rangle_{0\rm R}$ encode all information about the overlap
equations that the string coordinates have to satisfy. They can be determined respectively
\cite{Callan:1986bc,DiVecchia:1999uf} as 
\be\label{mmatrix}
M = ([(\eta -
\hat{F})(\eta + \hat{F})^{-1}]_{\alpha\beta},  -
\delta_{ij}),
\ee
\be\label{bzm}
|B_X\rangle_0 = [- \det
(\eta + \hat F)]^{1/2} \,\delta^{9 - p} (q^i - y^i) \prod_{\mu
= 0}^9 |k^\mu = 0\rangle,
\ee 
for the bosonic sector, and 
\be\label{rrzm}
|B_\psi, \eta\rangle_{0\rm R} = (C \Gamma^0 \Gamma^1
\cdots \Gamma^p \frac{1 + {\rm i} \eta \Gamma_{11}}{1 + {\rm i} \eta
} U )_{AB} |A\rangle |\tilde B\rangle,
\ee
 for the R-R sector. In
the above, the Greek indices $\alpha, \beta, \cdots$ label the
world-volume directions $0, 1, \cdots, p$ along which the Dp
brane extends, while the Latin indices $i, j, \cdots$ label the
directions transverse to the brane, i.e., $p + 1, \cdots, 9$. We define $\hat F = 2\pi \alpha' F$ with $F$ the external
worldvolume field. We also have denoted by $y^i$ the
positions of the D-brane along the transverse directions, by $C$ the
charge conjugation matrix and by $U$ the matrix 
\be\label{umatrix}
 U (\hat F) = \frac{1}{\sqrt{- \det (\eta + \hat F)}} ; {\rm exp} \left(- \frac{1}{2} {\hat
F}_{\alpha\beta}\Gamma^\alpha\Gamma^\beta\right); 
\ee
with the symbol $;\,\, ;$ denoting the indices of the $\Gamma$-matrices completely anti-symmetrized  in each term of the exponential expansion.
$|A \rangle |\tilde B\rangle$ stands for the spinor vacuum of the R-R
sector. Note that the $\eta$ in the above
denotes either sign $\pm$ or the worldvolume Minkowski flat metric and should be clear from the content.

 The vacuum amplitude can be calculated via
 \be\label{ampli}
 \Gamma = \langle B (\hat F') | D |B (\hat F) \rangle,
 \ee
  where $D$ is the closed string propagator defined as 
\be\label{prog}
 D =
\frac{\alpha'}{4 \pi} \int_{|z| \le 1} \frac{d^2 z}{|z|^2} z^{L_0}
 {\bar z}^{{\tilde L}_0}.
 \ee 
 Here $L_0$ and ${\tilde L}_0$ are
the respective left and right mover total zero-mode Virasoro
generators of matter fields, ghosts and superghosts. For example,
$L_0 = L^X_0 + L_0^\psi + L_0^{\rm gh} + L_0^{\rm sgh}$ where $L_0^X,
L_0^\psi, L_0^{\rm gh}$ and $L_0^{\rm sgh}$ represent contributions from
matter fields $X^\mu$, matter fields $\psi^\mu$, ghosts $b$ and $c$,
and superghosts $\beta$ and $\gamma$, respectively, and their
explicit expressions can be found in any standard discussion of
superstring theories, for example in \cite{Di Vecchia:1999rh},
therefore will not be presented here. The above total vacuum amplitude has
contributions from both NS-NS and R-R sectors, respectively, and can
be written as $\Gamma = \Gamma_{\rm NSNS} + \Gamma_{\rm RR}$. In
calculating either $\Gamma_{\rm NSNS}$ or $\Gamma_{\rm RR}$, we need to
keep in mind that the boundary state used should be the GSO
projected one as given earlier. For this purpose, we
need to calculate first the amplitude $ \Gamma (\eta',
\eta) = \langle B', \eta'| D |B, \eta\rangle $ in each sector
with $\eta' \eta = +\, {\rm or} - $, $B' = B (\hat F')$ and $B = B(\hat F)$. In doing so, we can set  $\tilde L_0 = L_0$ in the above propagator
due to the fact that $\tilde L_0 |B\rangle = L_0 |B\rangle$, which
can be used to simplify the calculations. Actually, $\Gamma(\eta', \eta)$ depends only on the product of $\eta'$ and $\eta$, i.e., $\Gamma(\eta', \eta) = \Gamma (\eta' \eta)$. In the NS-NS sector, this gives $\Gamma_{\rm NSNS} (\pm) \equiv \Gamma (\eta', \eta)$ when $\eta' \eta = \pm$, respectively.  Similarly we have $\Gamma_{\rm RR} (\pm) \equiv \Gamma (\eta', \eta)$ when $\eta'\eta = \pm$ in the R-R sector. We then have
\be\label{ns-r}
\Gamma_{\rm NSNS} = \frac{1}{2} \left[\Gamma_{\rm NSNS} (+) - \Gamma_{\rm NSNS} (-)\right], \quad \Gamma_{\rm RR} = \frac{1}{2}\left[\Gamma_{\rm RR} (+) + \Gamma_{\rm RR} (-) \right].
\ee
 Given the structure of the
boundary state, the amplitude $\Gamma (\eta'
\eta)$ can be factorized as 
\be\label{amplitude}
\Gamma (\eta' \eta) = \frac{
n_1 n_2 c_p^2}{4} \frac{\alpha'}{4 \pi} \int_{|z| \le 1} \frac{d^2 z}{|z|^2}
A^X \, A^{\rm bc}\, A^\psi (\eta'\eta)\, A^{\beta\gamma} (\eta'
\eta),\ee where we have replaced the $c_p$ in the boundary state
by $n c_p$ with $n$ an integer to count the
multiplicity of  Dp branes. In the above, we have 
\bea\label{me}
&&A^X = \langle B'_X | |z|^{2 L^X_0} |B_X \rangle,\quad
A^\psi (\eta' \eta) = \langle B'_\psi, \eta'| |z|^{2 L_0^\psi}
|B_\psi, \eta \rangle, \nn
&&A^{\rm bc} = \langle B_{\rm gh} | |z|^{2
L_0^{\rm gh}} | B_{\rm gh}\rangle,\quad A^{\beta\gamma} (\eta' \eta) =
\langle B_{\rm sgh}, \eta'| |z|^{2 L_0^{\rm sgh}} |B_{\rm sgh}, \eta\rangle.
\eea
The above ghost and superghost matrix elements $A^{\rm bc} $ and $A^{\beta\gamma} (\eta' \eta)$, both independent of the fluxes, can be calculated to give,
\be\label{ghme}
A^{\rm bc} = |z|^{- 2} \prod_{n = 1}^\infty \left(1 - |z|^{2n}\right)^2,
\ee 
and in the NS-NS sector
\be\label{sghmensns}
A_{\rm NSNS}^{\beta\gamma} (\eta' \eta) = |z| \prod_{n = 1}^\infty \left(1 + \eta'\eta |z|^{2n - 1}\right)^{-2},
\ee
while in the R-R sector
\be\label{sghmerr}
 A_{\rm RR}^{\beta\gamma} (\eta' \eta) =  {}_{\rm R0}\langle B_{\rm sgh}, \eta'|B_{\rm sgh},\eta\rangle_{\rm 0R}\, |z|^{\frac{3}{4}} \prod_{n = 1}^\infty \left(1 + \eta'\eta |z|^{2n}\right)^{-2},
\ee
where ${}_{\rm R0}\langle B_{\rm sgh}, \eta'|B_{\rm sgh},\eta\rangle_{\rm 0R}$ denotes the superghost zero-mode contribution which requires a regularization along with the zero-mode contribution of matter field $\psi$ in this sector.  We will discuss this regularization later on. 
   
    For the matrix elements of matter part, i.e. $A^X$ and $A^\psi (\eta'\eta)$ given in \eqref{me},  we can also calculate them with the matrix $M$ given in \eqref{mmatrix}. Their computations can be greatly simplified if the following property of matrix $M$ is used,
 \be\label{matrixmp}
 M_\mu\,^\rho (M^T)_\rho\,^\nu = (M^T)_\mu\,^\rho M_\rho\,^\nu = \delta_\mu\,^\nu,
 \ee
 where $T$ denotes the transpose of matrix.  For a system of two sets of Dp branes, placed parallel at a separation $y$, with one carrying flux $\hat F'$ and the other carrying flux $\hat F$, we can then have, 
 \be\label{matrixx}
 A^X = V_{p + 1} \frac{\left[\det(\eta + \hat F') \det(\eta + \hat F)\right]^{\frac{1}{2}}} { \left(2 \pi^2 \alpha' t\right)^{ \frac{9 - p}{2}}}\, e^{- \frac{y^2}{2\pi \alpha' t}} \prod_{n = 1}^\infty \left(\frac{1}{1 - |z|^{2n}}\right)^{9 - p} \prod_{\alpha = 0}^p \frac{1}{1 - \lambda_\alpha |z|^{2n}},
 \ee
and in the NS-NS sector
\be\label{matrixpsinsns}
A^\psi_{\rm NSNS} (\eta'\eta) = \prod_{n = 1}^\infty \left(1 + \eta'\eta |z|^{2n - 1}\right)^{9 - p} \prod_{\alpha = 0}^p \left(1 + \eta'\eta \lambda_\alpha |z|^{2n - 1}\right),
\ee
while in the R-R sector
\be\label{matrixpsirr} 
A^\psi_{\rm RR} (\eta'\eta) = {}_{R0}\langle B'_\psi, \eta' |B_\psi, \eta\rangle_{\rm 0R}\, |z|^{\frac{5}{4}} \prod_{n = 1}^\infty \left(1 + \eta'\eta |z|^{2n}\right)^{9 - p} \prod_{\alpha =0}^p \left(1 + \eta' \eta \lambda_\alpha |z|^{2n}\right),
\ee  
 where $ {}_{R0}\langle B'_\psi, \eta' |B_\psi, \eta\rangle_{\rm 0R}$ denotes the zero-mode contribution in this sector mentioned earlier.  In the above, $|z| = e^{- \pi t}$, $V_{p + 1}$ denotes the volume of the Dp brane worldvolume, $\lambda_\alpha$ are the eigenvalues of the matrix $w_{(1 + p) \times (1 + p)}$ defined as
 \be\label{matrixw}
 W = M M'^T = \left(\begin{array}{cc}
 w_{(1 + p)\times(1 + p)} & 0\\
 0 & \mathbb{I}_{(9 - p)\times (9 - p)}\end{array}\right).
 \ee
 where matrix $M'$ and $M$ are the one given in \eqref{mmatrix} when the corresponding fluxes are $\hat F'$ and $\hat F$, respectively, and $\mathbb{I}$ stands for the unit matrix.  The orthogonal matrix $W$, satisfying
 \be\label{orthogonalw}
 W W^T = W^T W = \mathbb{I}_{10 \times 10},
 \ee
 can be obtained from a redefinition of the certain oscillator modes, say $\tilde a_{n \nu}$, which is a trick used in simplifying the evaluation of the matrix elements of matter part from the contribution of oscillator modes.   Let us take the following as a simple illustration for obtaining the matrix $W$. In obtaining $A^X$,  we need to evaluate, for given $n > 0$, the following matrix element,
 \be\label{illustration}
 \langle 0| e^{- \frac{1}{n} \alpha_n^\mu (M')_\mu\,^\nu \tilde\alpha_{n}} |z|^{ 2\alpha_{- n}^\tau \alpha_{n \tau}} e^{- \frac{1}{n}  \alpha_{-n}^\rho (M)_\rho\,^\sigma \tilde\alpha_{-n\sigma} }| 0\rangle = \langle 0| e^{- \frac{1}{n} \alpha_n^\mu (M')_\mu\,^\nu \tilde\alpha_{n\nu}} e^{- \frac{|z|^{2n}}{\, \,n}  \alpha_{-n}^\rho (M)_\rho\,^\sigma \tilde\alpha_{-n\sigma} }| 0\rangle,
 \ee
 where $|0\rangle$ stands for the vacuum.  Purely for simplifying the evaluation of the matrix element on the right of the above equality, we first define $\tilde{\alpha'}_\mu = (M' )_{\mu}\,^{\rho}\tilde{\alpha}_\rho$ where we have omitted the index $n$ since this works for both $n > 0$ and $n < 0$, noting the matrix $M'$ being real.  Note that the commutation relation $[\tilde{\alpha'}_{n\,\mu}, \tilde{\alpha'}_{m\, \nu}] = \eta_{\mu\nu} \delta_{n + m, 0}$ continues to hold,  using the property of matrix $M'$ as given in \eqref{matrixmp}.  With this property of matrix $M'$, we can have $\tilde \alpha_{\mu} =  (M'^T)_{\mu}\,^{\nu} \tilde\alpha'_{\nu}$. Substituting this into \eqref{illustration} for $n < 0$ and also dropping the prime on $\tilde \alpha'$, we have \eqref{illustration} as
 \be\label{illustrationone}
 \langle 0| e^{- \frac{1}{n}  \tilde \alpha_{n}^{\mu}  \alpha_{n \mu} }e^{- \frac{|z|^{2n}}{\,\,n}  \alpha_{-n}^\rho W_\rho\,^\sigma \tilde\alpha_{-n\sigma}} | 0\rangle,
 \ee  
where $W$ is precisely the one given in \eqref{matrixw}.  Since $W$ is an unit matrix in the absence of fluxes, we expect that it can be diagonalized with the deformation of adding fluxes using the following non-singular matrix $V$,
\be\label{unitaryz}
V = \left(\begin{array}{cc}
v _{(1 + p)\times(1 + p)} & 0\\
  & \mathbb{I}_{(9 - p)\times (9 - p)}\end{array}\right),
 \ee
such that 
\be\label{diagonalw}
W = V W_0 V^{-1}.
\ee
In the above,
\be\label{w0}
W_0 = \left(\begin{array}{ccccc}
 \lambda_0 &&&&\\
 &\lambda_1&&&\\
 &&\ddots&&\\
 &&&\lambda_p&\\
 &&&&\mathbb{I}_{(9 - p)\times(9 - p)}
 \end{array}\right),
 \ee
and $v$ is a $(1 + p)\times (1 + p)$ non-singular matrix.  We further define\footnote{This purely serves the purpose of  simplifying  the evaluation of the matrix element \eqref{illustrationone}. For this, we keep the annihilation operator $\alpha'_{n \mu}$ with a lower Lorentz index $\mu$ while the creation operator $\alpha'^{\nu}_{-n}$ with an upper Lorentz index $\nu$.  It will be opposite for the corresponding oscillators with tilde.}, for $n > 0$, $\alpha'_{n\mu} =  (V^{-1})_{\mu}\,^{\nu}\alpha_{n \nu}$ and $\alpha'^{\mu}_{- n} =  \alpha^{\nu}_{-n}\, V_{\nu}\,^{\mu}$, and  $\tilde\alpha'_{-n\mu} =  (V^{-1})_{\mu}\,^{\nu}\tilde \alpha_{- n \nu}$ and $\tilde \alpha'^{\mu}_{n} =  \tilde \alpha^{\nu}_{n}\, V_{\nu}\,^{\mu}$. Note that now $ \tilde {\alpha'}^{\mu}_{n}\alpha'_{n \mu}=  \tilde\alpha^{\mu}_{n} \alpha_{n \mu}$. The matrix element  \eqref{illustrationone} becomes  
\be\label{illustwo}
\langle 0| e^{- \frac{1}{n}  \tilde \alpha'^{\mu}_{n} \alpha'_{n \mu} }e^{- \frac{|z|^{2n}}{\,\,n}  \lambda_{\rho}\, \alpha'^{\rho}_{-n} \tilde\alpha'_{-n\rho}} | 0\rangle.
\ee 
We have now the commutator relations $[\alpha'_{n \mu}, \alpha'^{\nu}_{- m}] = n \delta^{\nu}_{\mu} \delta_{n, m}$ and $[\tilde\alpha'^{\mu}_n, \tilde\alpha'_{- m \nu}]  = n \delta^\mu_\nu \delta_{n, m}$ when $n, m > 0$.  We now still have $\alpha'_{n \mu} | 0\rangle = \tilde\alpha'^{\mu}_{n} | 0\rangle = 0$ and $ \langle 0| \alpha'^{\mu}_{- n}  = \langle 0| \tilde\alpha'_{-n \mu} = 0$. The evaluation of \eqref{illustwo} becomes then as easy as the case without the presence of fluxes, giving the results of \eqref{matrixx} to \eqref{matrixpsirr}, respectively. 

We would like to point out that a similar approach to the above in simplifying the computations can also be adapted for a system of D$p'$ and D$p$, placed parallel at a separation, with each carrying fluxes for $p - p' = 2 k$ with $k = 0, 1 , 2, 3$.  What we have discussed above corresponds to $k = 0$ case.  

Given the general fluxes $\hat F'$ and $\hat F$, therefore $W$ from \eqref{matrixw} and \eqref{mmatrix}, what can we say about the eigenvalues $\lambda_\alpha$ of $W$ for $\alpha =0, 1, \cdots p$?  Since the matrix $W_\mu\,^\nu$ satisfies \eqref{orthogonalw} and it is an identical matrix when there is no flux present, so we must have
$\det W = 1$ and $\det w = 1$. Then from \eqref{diagonalw}, we have $\lambda_0 \lambda_1 \cdots \lambda_p = 1$,  the product of the eigenvalues being unity. If we take trace of matrix $W$, we end up $\sum_{\alpha = 0}^p \lambda_\alpha = {\rm tr} w$ from ${\rm Tr} W = \sum_{\alpha = 0}^p \lambda_\alpha  + (9 - p)$. Here we use the big trace symbol ${\rm Tr}$ denoting the trace of $W$ and the small trace symbol ${\rm tr}$ denoting the trace of $w$ defined in \eqref{matrixw}. Further we can have $\sum_{\alpha =0}^p \lambda^n_\alpha = {\rm tr w^n}$ with $n = 1, 2, \cdots $. We also have $W^{-1} = V W^{- 1}_0 V^{-1}$ and ${\rm Tr} (W^{- 1})^{n} = {\rm Tr} (W^T)^{n}$. This gives $\sum_{\alpha = 0}^p 1/\lambda^n_\alpha = \sum_{\alpha = 0}^p \lambda^n_\alpha$. Using these properties, one can show in general\footnote{The author would like to thank Zhihao Wu and Qiang Jia for discussion of this.}, using the eigenvalue equation $f (\lambda) = \det (\lambda \delta_{\alpha}^{\beta} - w_{\alpha}^{\beta}) = (\lambda - \lambda_{0}) (\lambda - \lambda_{1}) \cdots (\lambda - \lambda_{p}) = 0$, that if $p$ is even, one of the eigenvalues is unity and the remaining ones give $p/2$ pairs and for each pair the two eigenvalues are reciprocal to each other. For odd $p$, we have $(p + 1)/2$ pairs of eigenvalues and the two eigenvalues in each pair are also reciprocal to each other.  

With the above preparation, we are ready to give the general structure of $\Gamma_{\rm NSNS} (\eta'\eta)$ in the NS-NS sector and that of $\Gamma_{\rm RR} (\eta'\eta)$ in the R-R sector, respectively, for the system of two sets of Dp branes, placed parallel at a separation, carrying the general respective fluxes $\hat F'$ and $\hat F$.  For the NS-NS sector, using \eqref{amplitude},  \eqref{ghme}, \eqref{sghmensns}, \eqref{matrixx} and \eqref{matrixpsinsns},  we have then
\bea\label{amplitudensns}
\Gamma_{\rm NSNS} (\eta'\eta) &=& \frac{n_1 n_2 V_{p + 1} \sqrt{\det (\eta + \hat F')\det(\eta + \hat F)}}{(8 \pi^2 \alpha')^{\frac{1 + p}{2}}} \int_0^\infty \frac{d t} {t^{\frac{9 - p}{2}}} e^{- \frac{y^2}{2\pi\alpha' t}}\, |z|^{-1}\nn
&\,&\times\prod_{n = 1}^\infty \left(\frac{1  + \eta'\eta |z|^{2n - 1}}{1 - |z|^{2n}}\right)^{7 - p} \prod_{\alpha = 0}^p \frac{1 + \lambda_\alpha \eta'\eta |z|^{2 n - 1}}{1 - \lambda_\alpha |z|^{2n}},
\eea 
while in the R-R sector, we have from \eqref{amplitude}, \eqref{ghme}, \eqref{sghmerr}, \eqref{matrixx} and \eqref{matrixpsirr}, 
\bea\label{amplituderr}
\Gamma_{\rm RR} (\eta'\eta) &=& \frac{n_1 n_2 V_{p + 1} \sqrt{\det(\eta + \hat F')\det(\eta + \hat F)}}{(8\pi^2\alpha')^{\frac{1 + p}{2}}} {}_{\rm 0R}\langle B', \eta'| B, \eta\rangle_{\rm 0R}
\int_0^\infty \frac{dt}{t^{\frac{9 - p}{2}}} \, e^{- \frac{y^2}{2\pi \alpha' t}} \nn
&\,& \times \prod_{n = 1}^\infty \left(\frac{1 + \eta' \eta |z|^{2n}}{1 - |z|^{2n}}\right)^{7 - p} \prod_{\alpha = 0}^p \frac{1 + \eta'\eta \lambda_\alpha |z|^{2n}}{1 - \lambda_\alpha |z|^{2n}},
\eea
where the zero-mode contribution 
\be\label{0mme}
{}_{\rm 0R}\langle B', \eta'| B, \eta\rangle_{\rm 0R} \equiv {}_{\rm 0R}\langle B_{\rm sgh}, \eta'| B_{\rm sgh}, \eta\rangle_{\rm 0R} \times {}_{\rm 0R}\langle B'_\psi, \eta'| B_\psi, \eta\rangle_{\rm 0R} ,
\ee
whose respective explicit relation with the fluxes will be given in the specific cases considered in the following sections.  In obtaining the above,  we have used the following relations
\be\label{relations}
 \frac{c^2_p}{16 \pi (2 \pi^2 \alpha')^{\frac{7 - p}{2}} }= \frac{1}{(8\pi^2\alpha')^{\frac{1 + p}{2}}}, \qquad \int_{|z|\le 1}\frac{ d^2 z}{|z|^2} = 2\pi^2 \int_0^\infty dt 
 \ee
 where the explicit expression for $c_p$ as given right after \eqref{mgbs} has been used and $|z| = e^{- \pi t}$ as given earlier. 
  
So with the ready form of the amplitude (\ref{amplitudensns}) and (\ref{amplituderr}), the amplitude computations are  just boiled down to the determination of the eigenvalue $\lambda_{\alpha}$ and the evaluation of the zero-mode matrix (\ref{0mme}) once the worldvolume fluxes are given.     In the following three sections, we will compute the explicit interaction amplitude and analyze its analytical structure for each of the three cases given in the Introduction.  
 
\section{The $8 \ge p \ge 2$ case}
In this section, we will consider the following two subcases with the corresponding two non-vanishing field strength components on each set of Dp branes sharing a common index. Without loss of generality,  it can be cast in either of the following two structures 
\be\label{eestructure}
\hat F =\left( \begin{array}{ccccc}
0 & - f_1 &- f_2 &0 & \ldots\\
f_1&0&0&0&\ldots\\
f_2&0&0&0&\ldots\\
0&0&0&0&\ldots\\
 \vdots&\vdots&\vdots&\vdots&\ddots
 \end{array}\right)_{(1 + p)\times (1 + p)},
 \ee
  or
  \be\label{egstructure}
   \hat F =\left( \begin{array}{ccccc}
0 & - f & 0&0 & \ldots\\
f&0&- g&0&\ldots\\
0&g&0&0&\ldots\\
0&0&0&0&\ldots\\
 \vdots&\vdots&\vdots&\vdots&\ddots
 \end{array}\right)_{(1 + p)\times (1 + p)}.
  \ee 
In the first subcase, we have two electric fluxes $\hat F_{01} = - \hat F_{10} =  -f_1$ and $\hat F_{02} = - \hat F_{20} = - f_2$,  both of which share a common time index $`0'$ while in the second subcase, we have one electric flux $\hat F_{01} = - \hat F_{10} =  - f$ and a magnetic one $\hat F_{12} = - \hat F_{21} = - g$, both of which share a spatial index $`1'$. In what follows, let us consider each in order.
\subsection{The electric-electric case}
In this subsection, we first consider the $\hat F'$ and $\hat F$ with the structure given in \eqref{eestructure} to compute the interaction amplitude and subsequently to determine the open string pair production rate. 
\subsubsection{The interaction amplitude} 
Let us begin with computing the interaction amplitude. For this, we need to compute the corresponding $M'$ and $M$ via \eqref{mmatrix}, respectively, and then use \eqref{matrixw} to determine $w$.  From this $w$, we have the eigenvalues? following the description given in the previous section, as 
\bea\label{eeeigenv}
\lambda_0 \lambda_1 \lambda_2 &=& 1,\nn
\lambda_0 + \lambda_1 + \lambda_2 &=& \frac{1}{\lambda_0} + \frac{1}{\lambda_1} + \frac{1}{\lambda_2} =  \lambda_1 \lambda_2 + \lambda_0 \lambda_1 + \lambda_0 \lambda_2,\nn
& =& \frac{(1 + f^2_1 + f^2_2)(1 + f'^2_1 + f'^2_2) - 4 f_1 f'_1 - 4 f_2 f'_2}{(1 - f^2_1 - f^2_2)(1 - f'^2_1 - f'^2)}\nn
 &\,&    + \frac{(1 + f^2_1 - f^2_2)(1 + f'^2_1 - f'^2_2) - 4 f_1 f'_1 + 4 f_1 f_2 f'_1 f'_2}{(1 - f^2_1 - f^2_2)(1 - f'^2_1 - f'^2)}\nn
&\,& + \frac{(1 - f^2_1 + f^2_2)(1 - f'^2_1 + f'^2_2) - - 4 f_2 f'_2 + 4 f_1 f_2 f'_1 f'_2}{(1 - f^2_1 - f^2_2)(1 - f'^2_1 - f'^2)},
\eea
and the rest $\lambda_3 = \cdots = \lambda_p = 1$. In obtaining the last equality in the second line above, we have used the equation in the first line. We don't actually need to solve the eigenvalues $\lambda_0, \lambda_1, \lambda_2$ from \eqref{eeeigenv} and the relations satisfied by them are just needed to give the amplitude which we will compute now. 
 
 From \eqref{amplitudensns}, we have the NS-NS amplitude 
 \bea\label{amplitnsnsee}
 \Gamma_{\rm NSNS} (\eta'\eta) &=& \frac{n_1 n_2 V_{p + 1} \sqrt{ (1 - f^2_1 - f^2_2)(1 - f'^2_1 - f'^2_2)}}{(8 \pi^2 \alpha')^{\frac{1 + p}{2}}} \int_0^\infty \frac{d t} {t^{\frac{9 - p}{2}}}\, e^{- \frac{y^2}{2\pi\alpha' t}}\, |z|^{-1}\nn
&\,&\times\prod_{n = 1}^\infty \left(\frac{1  + \eta'\eta |z|^{2n - 1}}{1 - |z|^{2n}}\right)^5 \prod_{\alpha = 0}^2 \frac{1 + \lambda_\alpha \eta'\eta |z|^{2 n - 1}}{1 - \lambda_\alpha |z|^{2n}},
\eea 
where we can simplify the last product factor involving the eigenvalues $\lambda_0, \lambda_1$ and $\lambda_2$,  using the eigenvalue relations given in \eqref{eeeigenv}, as
\be\label{prods}
 \prod_{\alpha = 0}^2 \frac{1 + \lambda_\alpha \eta'\eta |z|^{2 n - 1}}{1 - \lambda_\alpha |z|^{2n}} = \frac{(1 + \eta'\eta |z|^{2n - 1}) (1 + \lambda \eta' \eta |z|^{2n - 1}) (1 + \lambda^{-1} \eta'\eta |z|^{2n - 1})} {(1 - |z|^{2n}) ( 1 - \lambda |z|^{2n}) (1 - \lambda^{-1} |z|^{2n})},
 \ee
where 
\bea\label{lambdaparameter}
\lambda + \lambda^{-1} &=& \lambda_0 + \lambda_1 + \lambda_2 - 1\nn
&=& 2 \frac{(f_1 - f'_1)^2 + (f_2 - f'_2)^2 + (1 - f_1 f'_1 - f_2 f'_2)^2 - (f_1 f'_2 - f_2 f'_1)^2}{(1 - f^2_1 - f^2_2)(1 - f'^2_1 - f'^2_2)}.
\eea
With this, we can re-express the above amplitude as
\bea\label{eeamplitnsns}
 \Gamma_{\rm NSNS} (\eta'\eta) &=& \frac{n_1 n_2 V_{p + 1} \sqrt{ (1 - f^2_1 - f^2_2)(1 - f'^2_1 - f'^2_2)}}{(8 \pi^2 \alpha')^{\frac{1 + p}{2}}} \int_0^\infty \frac{d t}{t^{\frac{9 - p}{2}}}\, e^{- \frac{y^2}{2\pi\alpha' t}}\, |z|^{-1}\nn
&\,&\times\prod_{n = 1}^\infty \left(\frac{1  + \eta'\eta |z|^{2n - 1}}{1 - |z|^{2n}}\right)^6  \frac{(1 + \lambda \eta'\eta |z|^{2 n - 1})(1 + \lambda^{-1} \eta'\eta |z|^{2 n - 1})}{(1 - \lambda |z|^{2n})(1 - \lambda^{-1} |z|^{2n})}.
\eea 
Then the total amplitude from the NS-NS sector is
\bea\label{eetotalamplitnsns}
\Gamma_{\rm NSNS} &=& \frac{1}{2} \left[ \Gamma_{\rm NSNS} ( +) - \Gamma_{\rm NSNS} (-)\right],\nn
&=&\frac{n_1 n_2 V_{p + 1} \sqrt{ (1 - f^2_1 - f^2_2)(1 - f'^2_1 - f'^2_2)}}{2(8 \pi^2 \alpha')^{\frac{1 + p}{2}}} \int_0^\infty \frac{d t} {t^{\frac{9 - p}{2}}}\, e^{- \frac{y^2}{2\pi\alpha' t}}\, |z|^{-1} \nn
&\,&\times\left[\prod_{n = 1}^\infty   A_n (+) - \prod_{n = 1}^\infty  A_n (-)\right],
\eea
where 
\be\label{A}
A_n (\eta\eta') = \left(\frac{1  + \eta\eta' |z|^{2n - 1}}{1 - |z|^{2n}}\right)^6\frac{(1 + \eta\eta'\, \lambda  |z|^{2 n - 1})(1 + \eta\eta'\, \lambda^{-1}  |z|^{2 n - 1})}{(1 - \lambda |z|^{2n})(1 - \lambda^{-1} |z|^{2n})}.
\ee
By the same token, we can have the R-R sector amplitude from \eqref{amplituderr} as
\bea\label{eeamplitrr}
\Gamma_{\rm RR} (\eta'\eta) &=& \frac{n_1 n_2 V_{p + 1} \sqrt{ (1 - f^2_1 - f^2_2)(1 - f'^2_1 - f'^2_2)}}{(8\pi^2\alpha')^{\frac{1 + p}{2}}} {}_{\rm 0R}\langle B', \eta'| B, \eta\rangle_{\rm 0R}
\int_0^\infty \frac{dt} {t^{\frac{9 - p}{2}}} \, e^{- \frac{y^2}{2\pi \alpha' t}} \nn
&\,& \times \prod_{n = 1}^\infty \left(\frac{1 + \eta' \eta |z|^{2n}}{1 - |z|^{2n}}\right)^6  \frac{(1 + \lambda \eta'\eta |z|^{2 n })(1 + \lambda^{-1} \eta'\eta |z|^{2 n})}{(1 - \lambda |z|^{2n})(1 - \lambda^{-1} |z|^{2n})}.
\eea
Following the regularization scheme given in \cite{Yost,Billo:1998vr}, we can have in the R-R sector, using the flux \eqref{eestructure} and the expression for the R-R sector zero-mode 
\eqref{rrzm} along with  \eqref{umatrix}, 
\be 
 {}_{\rm 0R}\langle B', \eta'| B, \eta\rangle_{\rm 0R}  =  - \frac{ 2^4 (1 - f_1 f'_1 - f_2 f'_2)}{\sqrt{(1 - f_1^2 - f^2_2)(1 - f'^2_1- f'^2_2) }} \delta_{\eta' \eta, +}.
 \ee
 So the total amplitude from the R-R sector is
 \bea\label{eetotalamplitrr}
 \Gamma_{\rm RR} &=& \frac{1}{2} \left[ \Gamma_{\rm RR} ( +) - \Gamma_{\rm RR} (-)\right],\nn
 &=&  - \frac{  2^3\, (1 - f_1 f'_1 - f_2 f'_2) n_1 n_2 V_{p + 1} }{(8\pi^2\alpha')^{\frac{1 + p}{2}}} 
\int_0^\infty \frac{ dt } {t^{\frac{9 - p}{2}}} \, e^{- \frac{y^2}{2\pi \alpha' t}} \,   \prod_{n = 1}^\infty B_n, \nn
\eea
where
  \be\label{B}
  B_n  =  \left(\frac{1 +  |z|^{2n}}{1 - |z|^{2n}}\right)^6  \frac{(1 + \lambda  |z|^{2 n })(1 + \lambda^{-1}  |z|^{2 n})}{(1 - \lambda |z|^{2n})(1 - \lambda^{-1} |z|^{2n})}.
 \ee 
  The total amplitude is then
 \bea\label{eetotalamplit}
 \Gamma &=& \Gamma_{\rm NSNS} + \Gamma_{\rm RR},\nn
 &=& \frac{n_1 n_2 V_{p + 1} \sqrt{ (1 - f^2_1 - f^2_2)(1 - f'^2_1 - f'^2_2)}}{2(8 \pi^2 \alpha')^{\frac{1 + p}{2}}} \int_0^\infty \frac{d t} { t^{\frac{9 -p}{2}}} e^{- \frac{y^2}{2\pi\alpha' t}} \left[ |z|^{-1} \left(\prod_{n = 1}^\infty A_n (+)\right.\right.\nn
 &\,& \left. \left. -  \prod_{n = 1}^\infty A_n (-)\right) -  \frac{ 2^4 (1 - f_1 f'_1 - f_2 f'_2)}{\sqrt{(1 - f_1^2 - f^2_2)(1 - f'^2_1- f'^2_2) }} \prod_{n = 1}^\infty B_n \right],
 \eea 
 where $A_n (\pm)$ are defined in \eqref{A} and $B_n$ in \eqref{B}. 

 Now we try to express this amplitude in terms of  the Dedekind $\eta$-function and various $\theta$-functions with their
standard definitions as given, for example, in \cite{polbookone} .   For this, we set the parameter $\lambda = e^{2\pi i \nu}$.  We  have then from \eqref{lambdaparameter}
 \bea\label{nuparameter}
 \cos \pi \nu &=& \frac{1 - f_1 f'_1 - f_2 f'_2}{\sqrt{(1 - f^2_1- f^2_2)(1 - f'^2_1 - f'^2_2)}}, \nn
  \sin\pi \nu &=&  \frac{\sqrt{(f_1 f'_2 - f_2 f'_1)^2 - (f_1 - f'_1)^2 - (f_2 - f'_2)^2}}{\sqrt{(1 - f^2_1 - f^2_2)(1 - f'^2_1 - f'^2_2)}}.
\eea
With this,  the total amplitude \eqref{eetotalamplit} can be expressed as 
\bea\label{ee-totalamplit}
 \Gamma &=& \Gamma_{\rm NSNS} + \Gamma_{\rm RR},\nn
 &=& \frac{n_1 n_2 V_{p + 1} \left[(f_1 f'_2 - f_2 f'_1)^2 - (f_1 - f'_1)^2 - (f_2 - f'_2)^2 \right]^{\frac{1}{2}}}{(8 \pi^2 \alpha')^{\frac{1 + p}{2}}} \int_0^\infty \frac{d t}{ t^{\frac{9 - p}{2}}} e^{- \frac{y^2}{2\pi\alpha' t}} \nn
 &\,& \times \frac{\theta_3(\nu| it) \theta^3_3 (0 | it) - \theta_4 (\nu | it) \theta^3_4 (0 | it) - \theta_2 (\nu | it) \theta^3_2 (0 | it)} {\eta^9 (it) \theta_1 (\nu | it)},\nn
 &=& \frac{2 n_1 n_2 V_{p + 1} \left[(f_1 f'_2 - f_2 f'_1)^2 - (f_1 - f'_1)^2 - (f_2 - f'_2)^2\right]^{\frac{1}{2}}}{(8 \pi^2 \alpha')^{\frac{1 + p}{2}}} \int_0^\infty \frac{d t}{t^{\frac{9 - p}{2}}}\, \frac{e^{- \frac{y^2}{2\pi\alpha' t}}\,\theta^4_1 (\frac{\nu}{2} | it)}{\eta^9 (it)\,\theta_1 (\nu | it)},\nn
  \eea 
where in the last equality  the following identity has been used
\be\label{theta-identity}
 2 \, \theta^4_1 \left(\left. \frac{\nu}{2}\right| \tau\right)  = \theta^3_3 (0 | \tau) \theta_3 (\nu|\tau)  - \theta^3_4 (0 | \tau) \theta_4(\nu |\tau)  - \theta_2^3 (0 |\tau) \theta_2 ( \nu | \tau), 
\ee
which is a special case of more general identity given in \cite{whittaker-watson}. Note that one can  show  $1 - f_1 f'_1 - f_2 f'_2 >  \sqrt{(1 - f^2_1- f^2_2)(1 - f'^2_1 - f'^2_2)}$, given that $f^2_1 + f_2^2 < 1$ and $f'^2_1 + f'^2_2  < 1$, which implies that the parameter $\nu$ is actually imaginary. So we can set $\nu = i \nu_0$ with $0 < \nu_0 < \infty$ and \eqref{nuparameter} becomes 
  \bea\label{nu0parameter}
 \cosh \pi \nu_0 &=& \frac{1 - f_1 f'_1 - f_2 f'_2}{\sqrt{(1 - f^2_1- f^2_2)(1 - f'^2_1 - f'^2_2)}}, \nn
  \sinh\pi \nu_0 &=&  \frac{\sqrt{ (f_1 - f'_1)^2 + (f_2 - f'_2)^2 - (f_1 f'_2 - f_2 f'_1)^2}}{\sqrt{(1 - f^2_1 - f^2_2)(1 - f'^2_1 - f'^2_2)}}.
\eea 
 From the last equality in \eqref{ee-totalamplit}, we have 
 \bea\label{eetotalamplit-explicit}
 \Gamma &=& \frac{ 2^4 n_1 n_2 V_{p + 1} \left[(1 - f^2_1 - f^2_2)(1 - f'^2_1 - f'^2_2)\right]^{\frac{1}{2}} \sinh^4 \frac{\pi \nu_0}{2}}{(8 \pi^2 \alpha')^{\frac{1 + p}{2}}} \int_0^\infty \frac{d t}{t}\,  t^{\frac{p -7}{2}} e^{- \frac{y^2}{2\pi\alpha' t}}\nn
 &\,&\times \prod_{n = 1}^\infty \frac{\left(1 - 2 |z|^{2n} \cosh\pi \nu_0 + |z|^{4n}\right)^4}{(1 - |z|^{2n})^6 (1 - 2 |z|^{2n} \cosh 2 \pi \nu_0 + |z|^{4n})}.
 \eea 
 For large $y$, the major contribution to the amplitude  is from the large $t$-integration and we have, for $p < 7$,
 \be
 \Gamma ({\rm large} \, y) \approx  n_1 n_2 V_{p + 1} \left[(1 - f^2_1 - f^2_2)(1 - f'^2_1 - f'^2_2)\right]^{\frac{1}{2}} \sinh^4 \frac{\pi \nu_0}{2} \, \frac{4 \pi (4 \pi^2 \alpha')^{3 - p}}{( 7  - p) \Omega_{8 - p}} \frac{1}{y^{7 - p}} > 0,
 \ee 
which gives indeed an attractive force\footnote{Our convention is that $\Gamma > 0$ gives an attractive force.}  as anticipated in the Introduction. Here $\Omega_q$ denotes the volume of unit q-sphere.  For small $y$, the small $t$ integration becomes important. The only factor which can become negative at small $t$ is the one $(1 - 2 |z|^{2n} \cosh 2 \pi \nu_0 + |z|^{4n})$ in the denominator of the infinite product in the integrand in  \eqref{eetotalamplit-explicit} since now $|z| \sim 1$ and $\cosh2\pi \nu_0 > 1$.  When this factor becomes negative, the sign for the infinite product remains unclear. So it is unclear about the nature of the interaction at small brane separation in terms of the closed string cylinder variable $t$.
\subsubsection{The open string pair production} 
For small $y$, the open string description is more suitable and the underlying physics becomes more clear. So let us now pass from the above closed string cylinder amplitude to the open string annulus one via the Jacobi transformation $t \to t' = 1/t$.  For this, we need the following relations for the Dedekind $\eta$-function and the $\theta_1$-function,
\be\label{jacobi}
\eta (\tau) = \frac{1}{(- i \tau)^{1/2}} \eta \left(- \frac{1}{\tau}\right), \quad \theta_1 (\nu | \tau) = i \frac{e^{- i \pi \nu^2/\tau}}{(- i \tau)^{1/2}} \theta_1 \left(\left.\frac{\nu}{\tau} \right| - \frac{1}{\tau}\right).
\ee 
Using these two relations with $\tau = it$ and $t' = 1/t$, we have from the last equality of \eqref{eetotalamplit} 
\bea\label{eeannulusamplit}
\Gamma &=&  \frac{2\, n_1 n_2 V_{p + 1} \left[ (f_1 - f'_1)^2 + (f_2 - f'_2)^2 - (f_1 f'_2 - f_2 f'_1)^2\right]^{\frac{1}{2}}}{(8 \pi^2 \alpha')^{\frac{1 + p}{2}}} \int_0^\infty \frac{d t'}{t'^{\frac{1+ p}{2}}}\, \frac{e^{- \frac{y^2 t'}{2\pi\alpha' }}}{\eta^9 (it')} \frac{\theta^4_1 (\frac{\nu_0 t'}{2} | it')}{\theta_1 (\nu_0 t' | it')},\nn
&=& \frac{2^4 \,n_1 n_2 V_{p + 1} \left[ (f_1 - f'_1)^2 + (f_2 - f'_2)^2 - (f_1 f'_2 - f_2 f'_1)^2\right]^{\frac{1}{2}}}{(8 \pi^2 \alpha')^{\frac{1 + p}{2}}} \int_0^\infty \frac{d t}{t^{ \frac{1+ p}{2}}}\, e^{- \frac{y^2 t}{2\pi\alpha'}} \, \frac{\sin^4 \frac{\pi \nu_0 t}{2}}{\sin\pi \nu_0 t} \nn
&\,& \times\,  \prod_{n =1}^\infty \frac{(1 - 2 |z|^{2n} \cos \pi \nu_0 t + |z|^{4n})^4}{(1 - |z|^{2n})^6 (1 - 2 |z|^{2n} \cos2\pi \nu_0 t + |z|^{4n})},
\eea  
 where in the first equality we have set $\nu = i \nu_0$ and in the second equality we have dropped the prime on $t$ and  expressed out the Dedekind $\eta$-function and the 
 $\theta_1$-functions.  Here $|z|$ is the same as before, i.e. $|z| = e^{ - \pi t} < 1$.
 
   Note that from either \eqref{eetotalamplit-explicit} or \eqref{eeannulusamplit}, we can check that $\Gamma = 0$ only if $\nu_0 = 0$, which actually implies $f_1 = f'_1,  f_2 = f'_2$. In other words, the underlying system is still a 1/2 BPS one just like each set of the Dp branes carrying the same two electric fluxes\footnote{\label{fn7}The Dp branes carrying two orthogonal electric fluxes correspond to the so-called 1/2 BPS non-threshold  (F, F, Dp) bound state which can be obtained from the 1/2 BPS non-threshold bound state $D1\perp D1 \perp D1$  given in \cite{Lu:1999uv} using  T-dualities and S-duality.}. When $\Gamma \neq 0$, the integrand in \eqref{eeannulusamplit} has no exponential growing factor for large $t$, therefore there is no open string tachyon mode to appear for the present case of only having electric fluxes present.   For small $y$, we see that  all the other factors in the integrand in the second equality of \eqref{eeannulusamplit} are positive except for the factor $\sin\pi \nu_0 t$ in the denominator which oscillates between $+1$ and $-1$ as the variable $t$ increases. So this makes the small separation interaction nature obscure but also interesting. It is precisely due to the presence of this factor, which gives rise to an infinity
number of simple poles along the positive $t$-axis  in the integrand,  that signals a new physics process to occur.  These simple poles happen when $\sin\pi \nu_0 t$ vanishes while the factor $\sin \pi \nu_0 t /2$ does not.  We therefore have them at
\be\label{eesimplep}
t_k = \frac{2 k + 1}{\nu_0}, \qquad k = 0, 1,  \cdots.
\ee
Each of these simple poles actually tells the production of a pair of open strings 
as described in the Introduction under the action of the applied electric fluxes \cite{Bachas:1992bh,Bachas:1995kx}, whose masses are proportional to the brane separation.  When the brane separation is large, the probability in producing this kind of open string pairs is small since the mass for each pair is also large and therefore it is difficult to produce them. In this sense, the underlying system has almost no decay process to occur and the only thing left is the interaction between the two set of Dp branes.  We are therefore certain about the nature of the interaction. However, for small brane separation, the pair production can become significant and the system decays. In other words, the amplitude has actually an imaginary part reflecting this decay and also giving the pair production rate.  The infinite number of simple poles appearing in the integrand in \eqref{eeannulusamplit} indicates the occurrence of the open string pair production. This process will continue  and the energy of the system will be carried away by the pair production until $f_1 = f_1'$ and $f_2 = f'_2$ for which the system reaches its stable 1/2 BPS one. 

The rate of open string pair production per unit worldvolume is the imaginary
part of the amplitude \eqref{eeannulusamplit} in the second equality, which can be obtained as the sum of the residues of the poles of the integrand  times $\pi$ following \cite{Bachas:1992bh,Bachas:1995kx} and is given as
\bea\label{ee-ospr}
{\cal W} &=& - \frac{2 \,{\rm Im}\Gamma}{V_{p + 1}},\nn
&=&\frac{2^5 \,n_1 n_2  \left[ (f_1 - f'_1)^2 + (f_2 - f'_2)^2 - (f_1 f'_2 - f_2 f'_1)^2\right]^{\frac{1}{2}}}{ \nu_0\, (8 \pi^2 \alpha')^{\frac{1 + p}{2}}} \sum_{k = 0}^\infty \left(\frac{\nu_0}{2 k + 1}\right)^{\frac{p + 1}{2}} \, e^{ - \frac{(2 k + 1) y^2}{2\pi \nu_0 \alpha'}}  \nn
&\,& \times \prod_{n = 1}^\infty \left(\frac{1 + e^{- \frac{2 n (2k + 1) \pi}{\nu_0}}}{1 - e^{- \frac{2 n (2k + 1) \pi}{\nu_0}}} \right)^8.
 \eea
 We here try to understand this open string pair production rate.  As anticipated, for given $k$, the larger the brane separation is, the smaller the rate.  For given $y$, the larger the $k$ is, the smaller the rate, too.  This can be understood as for large $k$, the open string is produced with a tension $(2k + 1)$ times the fundamental string tension,  having also a large mass, therefore more difficult to be produced.  For given $k$ and $y$,  the larger the $\nu_0$ is, which can also imply a larger factor $\left[ (f_1 - f'_1)^2 + (f_2 - f'_2)^2 - (f_1 f'_2 - f_2 f'_1)^2\right]^{\frac{1}{2}}$ from \eqref{nu0parameter}, the larger the rate, too.  In particular, when $|f_a|$ and $|f'_a|$ with $a = 1, 2$ are given, $f_a f'_a < 0$, i.e., when the two electric fluxes on one set of Dp branes are opposite in directions to their correspondences on the other set, will give the largest $\nu_0$.  When either $\sqrt{f^2_1 + f^2_2} \to 1$ or
 $\sqrt{f'^2_1 + f'^2_2} \to 1$ or both, i.e., to their respective critical values, $\nu_0 \to \infty$ and the rate ${\cal W}$ blows up,  the onset of pair production instability.  
 
 For very small $\nu_0 \ll 1$,  the rate \eqref{ee-ospr} can be approximated by the first $k = 0$ term as
 \be\label{smallnu0-rate}
 {\cal W} \approx \frac{2^5 \,n_1 n_2 \sqrt{(1 - f^2_1 -f^2_2)(1 - f'^2_1- f'^2_2)}}{(8 \pi^2 \alpha')^{\frac{1 + p}{2}}}\, \nu_0^{\frac{p + 1}{2}} \, e^{ - \frac{ y^2}{2\pi \nu_0 \alpha'}} ,
 \ee
where we have used \eqref{nu0parameter} for $\nu_0 \ll 1$.  This rate is vanishing small and has no practical physical significance.  Note that when $f_a = f'_a$ (now $\nu_0 = 0$), the rate ${\cal W}$ vanishes and this is consistent with the underlying system being a 1/2 BPS stable one. 

    Before we close this subsection, we would like to point out that the present open string production \eqref{ee-ospr} results from the virtual open string pair connecting the two sets of Dp branes placed parallel at a separation under the action of the applied two electric fluxes on each set.  This is different from the open string pair production discussed in \cite{Bachas:1992bh,Bachas:1995kx} for which the virtual open string pair with their ends attaching on the same branes,  either on D26 branes in the bosonic case or on the D9 branes in Type I case.   If we really want to draw the analog, their rate is the one on each isolated set of Dp branes carrying electric flux(es).  For the present case, this rate actually vanishes since the open strings are neutral ones. This is also consist with the fact that each set of Dp branes carrying the electric fluxes is actually 1/2 BPS stable non-threshold bound state (see footnote \ref{fn7}) and therefore there should be no open string pair production. Note that the open string pair production ${\cal W}$ \eqref{ee-ospr} also vanishes when $f_a = f'_a$ (now $\nu_0 = 0$) and this is consistent with the underlying system being a 1/2 BPS stable one, too.  Further, unlike the amplitude and the open string pair production given in \cite{Bachas:1992bh,Bachas:1995kx} or in \cite{Lu:2009yx} for which the electric flux(es) is (are) along the same or opposite direction, the present ones are for the electric fluxes in different direction, for example, the total electric flux in one set of Dp branes has in general different magnitude and direction from that in the other set of Dp branes.  So the results in   \cite{Lu:2009yx} is just a special case of the present ones when we take, say, $f_{2} = f'_{2} = 0$.

\subsection{The electric-magnetic case}
We now repeat the same process as in the previous subsection but having the  $\hat F'$ and $\hat F$ with the structure given in \eqref{egstructure}. As we will see, this case\footnote{\label{fn8} Note that each such Dp is also 1/2 BPS non-threshold bound state which can be obtained from, say, 1/2 BPS non-threshold bound state ((F, D0), D2) given in \cite{Lu:1999uv}, by T-dualities along directions transverse to this bound state. Here the F-string is along one of D2 directions.}  is more richer in physics and has actually three subcases to consider.  In particular, we find an open string pair production enhancement which does not appear in the one-flux case considered previously by the present author and his colloaborator in \cite{Lu:2009au}.

\subsubsection{The interaction amplitude}
So we have the flux $\hat F'$ on one set of the Dp branes  and the flux $\hat F$ on the other set for the present subcase, respectively, as
\be\label{egflux}
   \hat F' =\left( \begin{array}{ccccc}
0 & - f' & 0&0 & \ldots\\
f'&0&- g'&0&\ldots\\
0&g'&0&0&\ldots\\
0&0&0&0&\ldots\\
 \vdots&\vdots&\vdots&\vdots&\ddots
 \end{array}\right)_{(1 + p)\times (1 + p)}, \quad 
   \hat F =\left( \begin{array}{ccccc}
0 & - f & 0&0 & \ldots\\
f&0&- g&0&\ldots\\
0&g&0&0&\ldots\\
0&0&0&0&\ldots\\
 \vdots&\vdots&\vdots&\vdots&\ddots
 \end{array}\right)_{(1 + p)\times (1 + p)}.
   \ee 

Following the same steps as in the previous subsection, we have the eigenvalues 
\bea\label{eg-eigenv}
\lambda_0 \lambda_1 \lambda_2 &=& 1, \nn
\lambda_0 + \lambda_1 + \lambda_2 &=& \frac{1}{\lambda_0} + \frac{1}{\lambda_1} + \frac{1}{\lambda_2} = \lambda_1 \lambda_2 + \lambda_0 \lambda_2 + \lambda_0 \lambda_1 \nn
                                                            &=& \frac{(1 + f^2 + g^2)(1 + f'^2 + g'^2) + (1 + f^2 - g^2)(1 + f'^2 - g'^2) }{(1 - f^2 + g^2)(1 - f'^2 + g'^2)} \nn
                                                            &\,& + \frac{ (1 - f^2 - g^2)(1 - f'^2 - g'^2) - 8 f f' +  8 g g' - 8 f f' g g'}{(1 - f^2 + g^2)(1 - f'^2 + g'^2)}. 
\eea
and $\lambda_3 = \cdots =\lambda_p = 1$.  The zero-mode contribution \eqref{0mme}  in the R-R sector to the amplitude \eqref{amplituderr} in the present case can be evaluated as
\be\label{eg0mme}
{}_{\rm 0R}\langle B', \eta'| B, \eta\rangle_{\rm 0R}  = - \frac{ 2^4 (1 - f f' + g g')}{\sqrt{(1 - f^2 + g^2)(1 - f'^2 + g'^2) }} \delta_{\eta' \eta, +}.
\ee
Using \eqref{amplitudensns} and \eqref{amplituderr} as well as \eqref{ns-r}, we can have the present total tree-level closed string cylinder interaction amplitude as
\bea\label{egtotalamplit}
 \Gamma &=& \Gamma_{\rm NSNS} + \Gamma_{\rm RR},\nn
 &=& \frac{n_1 n_2 V_{p + 1} \sqrt{ (1 - f^2 + g^2)(1 - f'^2 + g'^2)}}{2(8 \pi^2 \alpha')^{\frac{1 + p}{2}}} \int_0^\infty \frac{d t} { t^{\frac{9 - p}{2}}} e^{- \frac{y^2}{2\pi\alpha' t}} \left[ |z|^{-1} \left(\prod_{n = 1}^\infty A_n (+)\right.\right. \nn
 &\,& \left. \left. -  \prod_{n = 1}^\infty A_n (-) \right) -  \frac{ 2^4 (1 - f f' + g g')}{\sqrt{(1 - f^2 + g^2)(1 - f'^2 + g'^2) }} \prod_{n = 1}^\infty B_n \right],
 \eea 
 where $A_n (\pm)$ and $B_n$ are also given by \eqref{A} and \eqref{B}, respectively,   but for now
 \bea\label{eg-lambda}
 \lambda + \lambda^{-1} &=& \lambda_0 + \lambda_1 + \lambda_2 - 1,\nn
                                        &=& 2 \frac{(f - f')^2 + (1 - f f' + g g')^2 + (g f' - f g')^2 - (g - g')^2}{(1 - f^2 + g^2)(1 - f'^2 + g'^2)}.
  \eea
 By setting $\lambda = e^{2\pi i \nu}$, we then have 
 \bea\label{eg-nuparameter}
 \cos\pi \nu &=& \frac{1 - f f' + g g'}{\sqrt{(1 - f^2 + g^2)(1 - f'^2 + g'^2)}}, \nn
  \sin\pi \nu &=& \frac{\sqrt{(g - g')^2  - (f - f')^2 - (g f' - f g')^2}}{\sqrt{(1 - f^2 + g^2)(1 - f'^2 + g'^2)}}.
 \eea
With this, the above amplitude \eqref{egtotalamplit} can also be cast in terms of $\theta$-functions and Dedekind $\eta$-function as
\bea\label{eg-totalamplit}
\Gamma &=& \frac{n_1 n_2 V_{p + 1} \sqrt{(g - g')^2 - (f - f')^2 - (g f' - f g')^2}}{(8 \pi^2 \alpha')^{\frac{1 + p}{2}}} \int_0^\infty \frac{d t} { t^{\frac{9 - p}{2}}} e^{- \frac{y^2}{2\pi\alpha' t}} \nn
&\,& \times \frac{\theta_3^3 (0| it) \theta_3 (\nu | it) - \theta^3_4 (0 | it) \theta_4 (\nu |it) - \theta^3_2 (0 | it) \theta_2 (\nu | it)}{\eta^9 (it)\, \theta_1 (\nu | it)},\nn
&=& \frac{2\, n_1 n_2 V_{p + 1} \sqrt{(g - g')^2 - (f - f')^2 - (g f' - f g')^2}}{(8 \pi^2 \alpha')^{\frac{1 + p}{2}}} \int_0^\infty \frac{d t}{  t^{\frac{9 -p}{2}}} \frac{e^{- \frac{y^2}{2\pi\alpha' t}} \, \theta^4_1 \left(\frac{\nu}{2}\right|\left. it\right)}{\eta^9 (it) \, \theta_1 (\nu | it)},\nn
&=& \frac{2^4\, n_1 n_2 V_{p + 1} \sqrt{(1 - f^2 + g^2)(1 - f'^2 + g'^2)} \, \sin^4 \frac{\pi \nu}{2}}{(8 \pi^2 \alpha')^{\frac{1 + p}{2}}} \int_0^\infty \frac{d t} { t^{\frac{9 - p}{2}}} e^{- \frac{y^2}{2\pi\alpha' t}} \nn
&\,& \times\prod_{n = 1}^\infty  \frac{\left[1 - 2 |z|^{2n} \cos \pi \nu + |z|^{4n}\right]^4 }{(1 - |z|^{2n})^6 \left[1 - 2 |z|^{2n} \cos 2\pi \nu + |z|^{4n}\right]},
\eea
where in the second equality we have also used the identity \eqref{theta-identity} for various $\theta$-functions,  in the last equality we have used the explicit expressions for the 
Dedekind $\eta$-function and the $\theta_1$-function and again $|z| = e^{ - \pi t} < 1$.

     For large $y$,  the main contribution to the amplitude comes again from the large $t$ integration for which the infinite product can be approximated by $1$.  Then the integration can be carried out and is finite for $p < 7$ as in the previous case.  This gives  $\Gamma \propto 1 /y^{7 - p} > 0$,  an attractive interaction as expected.  For further analysis, we need to consider the following three subcases:  1)  $( g - g')^2 \ge (f - f')^2 + (g f' - f g')^2$; 2) $(g - g')^2 < (f - f')^2 + (g f' - f g')^2$ and $1 - f f' + g g' \ge 0$; 3)  $(g - g')^2 < (f - f')^2 + (g f' - f g')^2$ and $1 - f f' + g g' < 0$. We now consider each of these subcases in order.

\noindent     
{\bf Subcase 1):} For this, we have from \eqref{eg-nuparameter} that $\nu = \nu'_0$ is real and falls in the range of $0 < \nu'_0 < 1$.  Also from this equation and from the last equality in \eqref{eg-totalamplit}, we have that $\Gamma = 0$ gives $\nu = \nu'_0 = 0$, implying  $1 - f f' + g g'  = \sqrt{(1 - f^2 + g^2) (1 - f'^2 + g'^2)} > 0$. For $\nu'_0 \neq 0$, every factor in the integrand in the last equality of \eqref{eg-totalamplit} is positive, therefore the interaction is attractive, resembling a pure magnetic case. For this reason, we expect to see an open string tachyon to appear at small brane separation. For this, we need to re-express the amplitude as the open string annulus one via the Jacobi transformation $t \to t' = 1/t$. Using \eqref{jacobi}, the open string annulus amplitude can be obtained from the second equality of \eqref{eg-totalamplit} as
\bea\label{eg-annulusamplit}
\Gamma &=& \frac{2\, n_1 n_2 V_{p + 1} \sqrt{(g - g')^2 - (f - f')^2 - (g f' - f g')^2}}{(8 \pi^2 \alpha')^{\frac{1 + p}{2}}} \int_0^\infty \frac{d t'}{  t'^{\frac{1 + p}{2}}} \frac{e^{- \frac{y^2 t'}{2\pi\alpha' }}\, \theta^4_1 \left(\frac{ - i \nu'_0 t'}{2}\right|\left. it' \right)}{\eta^9 (it') \, \theta_1 (- i \nu'_0 t' | i t')},\nn
&=& \frac{2^4\, n_1 n_2 V_{p + 1} \sqrt{(g - g')^2 - (f - f')^2 - (g f' - f g')^2}}{(8 \pi^2 \alpha')^{\frac{1 + p}{2}}} \int_0^\infty \frac{d t}{  t^{\frac{1 + p}{2}}} e^{- \frac{y^2 t}{2\pi\alpha' }} \, \frac{\sinh^4 \frac{\pi \nu'_0 t}{2}}{\sinh \pi \nu'_0 t},\nn
&\,& \times \prod_{n = 1}^\infty \frac{\left(1 - 2 |z|^{2n} \cosh\pi\nu'_0 t + |z|^{4n}\right)^4}{(1 - |z|^{2n})^6 \left(1 - 2 |z|^{2n} \cosh 2\pi \nu'_0 t + |z|^{4n}\right)},
\eea
where in the second equality we have dropped the prime on $t$ and again $|z| = e^{- \pi t} < 1$. Each factor in the above integrand is also positive for $t > 0$, noting that $(1 - 2 |z|^{2n} \cosh2\pi \nu'_0 t + |z|^{4n}) =  (1 - e^{- 2\pi (n - \nu'_0) t}) (1 - e^{- 2\pi (n + \nu'_0) t}) > 0$ for $n \ge 1$. This again gives $\Gamma > 0$.  The integrand has no simple poles along the positive t-axis as expected.  For large $t$ (corresponding to small $y$),  we have 
\be 
\lim_{t \to \infty} \frac{\sinh^4 \frac{\pi \nu'_0 t}{2}}{\sinh \pi \nu'_0 t} \prod_{n = 1}^\infty \frac{\left(1 - 2 |z|^{2n} \cosh\pi\nu'_0 t + |z|^{4n}\right)^4}{(1 - |z|^{2n})^6 \left(1 - 2 |z|^{2n} \cosh 2\pi \nu'_0 t + |z|^{4n}\right)} \to e^{\pi \nu'_0 t} \to \infty,
\ee
implying the appearance of an open string tachyon mode as expected \cite{Banks:1995ch, Lu:2007kv}.  The tachyonic instability will be onset and the tachyon condensation will occur when $y \le \pi \sqrt{2 \nu \alpha'}$ \cite{Pesando:1999hm, Sen:1999xm}. 

\noindent 
{\bf Subcase 2):} For this case, we have $\nu = i \nu_0$ with $0 < \nu_0 < \infty$.  Now \eqref{eg-nuparameter} becomes 
\bea\label{eg-nu0parameter}
 \cosh\pi \nu_0 &=& \frac{1 - f f' + g g'}{\sqrt{(1 - f^2 + g^2)(1 - f'^2 + g'^2)}}, \nn
  \sinh\pi \nu_0 &=& \frac{\sqrt{ (f - f')^2 + (g f' - f g')^2 - (g - g')^2}}{\sqrt{(1 - f^2 + g^2)(1 - f'^2 + g'^2)}}.
 \eea
For this case, the effect of electric-fluxes dominates over that of magnetic ones.  The amplitude from \eqref{eg-totalamplit} is now
\bea\label{egnu0-totalamplit}
\Gamma &=& \frac{2\, i \, n_1 n_2 V_{p + 1} \sqrt{(f - f')^2 + (g f' - f g')^2 - (g - g')^2 }}{(8 \pi^2 \alpha')^{\frac{1 + p}{2}}} \int_0^\infty \frac{d t}{  t^{\frac{9 -p}{2}}} \frac{e^{- \frac{y^2}{2\pi\alpha' t}} \, \theta^4_1 \left(\frac{i\nu_0}{2}\right|\left. it\right)}{\eta^9 (it) \, \theta_1 (i\nu_0 | it)},\nn
&=& \frac{2^4\, n_1 n_2 V_{p + 1} \sqrt{(1 - f^2 + g^2)(1 - f'^2 + g'^2)} \, \sinh^4 \frac{\pi \nu_0}{2}}{(8 \pi^2 \alpha')^{\frac{1 + p}{2}}} \int_0^\infty \frac{d t} { t^{\frac{9 - 9}{2}}} e^{- \frac{y^2}{2\pi\alpha' t}} \nn
&\,& \times\prod_{n = 1}^\infty  \frac{\left[1 - 2 |z|^{2n} \cosh \pi \nu_0 + |z|^{4n}\right]^4 }{(1 - |z|^{2n})^6 \left[1 - 2 |z|^{2n} \cosh 2\pi \nu_0 + |z|^{4n}\right]}.
\eea
For large $y$, once again only large $t$-integration is important and this gives the finite amplitude $\Gamma \propto 1/ y^{7 - p} > 0$ for $p < 7$, which is attractive. For small $y$, the small $t$-integration can be important.  Then the factor $(1 - 2 |z|^{2n} \cosh 2 \pi \nu_0 + |z|^{4n})$ in the denominator of the infinite product in the  integrand can become negative and this makes the sign of the amplitude indefinte.  Our experience tells that this signals new physics, i.e., the open string pair production, to occur.  For this to be manifest, we need to pass the above tree-level closed string cylinder amplitude to the open string annulus one via the Jacobi transformation $t \to t' = 1/t$. For this, we also need to use the  
relations for the Dedekind $\eta$-function and the $\theta_1$-functions \eqref{jacobi}. We then have
\bea\label{egnu0-annulusamplit}
\Gamma &=& \frac{2 \,i\, n_1 n_2 V_{p + 1} \sqrt{(f - f')^2 + (g f' - f g')^2 - (g - g')^2}}{(8 \pi^2 \alpha')^{\frac{1 + p}{2}}} \int_0^\infty \frac{d t'}{  t'^{\frac{1 + p}{2}}} \frac{e^{- \frac{y^2 t'}{2\pi\alpha' }}\, \theta^4_1 \left(\frac{\nu_0 t'}{2}\right|\left. it' \right)}{\eta^9 (it') \, \theta_1 (\nu_0 t' | i t')},\nn
&=& \frac{2^4\, n_1 n_2 V_{p + 1} \sqrt{(f - f')^2 + (g f' - f g')^2 - (g - g')^2 }}{(8 \pi^2 \alpha')^{\frac{1 + p}{2}}} \int_0^\infty \frac{d t}{  t^{\frac{1 + p}{2}}} e^{- \frac{y^2 t}{2\pi\alpha' }} \, \frac{\sin^4 \frac{\pi \nu_0 t}{2}}{\sin \pi \nu_0 t},\nn
&\,& \times \prod_{n = 1}^\infty \frac{\left(1 - 2 |z|^{2n} \cos\pi\nu_0 t + |z|^{4n}\right)^4}{(1 - |z|^{2n})^6 \left(1 - 2 |z|^{2n} \cos 2\pi \nu_0 t + |z|^{4n}\right)},
\eea
where in the second equality we have dropped the prime on $t$ and here $|z| = e^{- \pi t} < 1$.  Apart from the overall factor $[(f - f')^2 + (g f' - f g')^2 - (g - g')^2]^{1/2}$,  the integrand in the second equality of \eqref{egnu0-annulusamplit} looks identical to that in \eqref{eeannulusamplit}.  So the physics is the same. For example, we have also
an infinite number of simple poles of the integrand occurring at $t_k = (2 k + 1)/\nu_0$ with $k = 0, 1, \cdots$ and the open string pair production rate is
\bea\label{eg-ospr}
{\cal W} &=& - \frac{2 \,{\rm Im}\Gamma}{V_{p + 1}},\nn
&=&\frac{2^5 \,n_1 n_2  \left[(f - f')^2 + (g f' - f g')^2 - (g - g')^2 \right]^{\frac{1}{2}}}{ \nu_0\, (8 \pi^2 \alpha')^{\frac{1 + p}{2}}} \sum_{k = 0}^\infty \left(\frac{\nu_0}{2 k + 1}\right)^{\frac{p + 1}{2}} \, e^{ - \frac{(2 k + 1) y^2}{2\pi \nu_0 \alpha'}}  \nn
&\,& \times \prod_{n = 1}^\infty \left(\frac{1 + e^{- \frac{2 n (2k + 1) \pi}{\nu_0}}}{1 - e^{- \frac{2 n (2k + 1) \pi}{\nu_0}}} \right)^8.
 \eea
For small $\nu_0 \ll 1$, the above rate can be approximated by the leading $k = 0$ term as
\be\label{egsmallnu0-w}
 {\cal W} \approx \frac{2^5 \,n_1 n_2 \sqrt{(1 - f^2 + g^2)(1 - f'^2 + g'^2)}}{(8 \pi^2 \alpha')^{\frac{1 + p}{2}}}\, \nu_0^{\frac{p + 1}{2}} \, e^{ - \frac{ y^2}{2\pi \nu_0 \alpha'}} ,
 \ee
where we have used \eqref{eg-nu0parameter} for small $\nu_0$. The only difference here from its counterpart in \eqref{smallnu0-rate} is that the magnetic fluxes appear to give some enhancement of this rate. The discussion about the production rate goes also the same as in the electric-electric case and will not repeat it here. 

\noindent
{\bf Subcase 3):}  This is the case which gives the open string pair production a significant enhancement and has not been seen previously, for example, as in the one-flux case \cite{Lu:2009au}.  For this case, $\nu = 1 - i \nu_0$ with $0 < \nu_0 < \infty$. We then have from \eqref{eg-nuparameter}
\bea\label{eg-nu02parameter}
 \cosh\pi \nu_0 &=& - \frac{1 - f f' + g g'}{\sqrt{(1 - f^2 + g^2)(1 - f'^2 + g'^2)}}, \nn
  \sinh\pi \nu_0&=& \frac{\sqrt{ (f - f')^2 + (g f' - f g')^2 - (g - g')^2 }}{\sqrt{(1 - f^2 + g^2)(1 - f'^2 + g'^2)}}.
 \eea
The amplitude \eqref{eg-totalamplit} in the second and third equalities, respectively,  becomes now
\bea\label{eg-nu02amplit}
\Gamma &=&  \frac{2\, i \, n_1 n_2 V_{p + 1} \sqrt{ (f - f')^2 +  (g f' - f g')^2 - (g - g')^2 }}{(8 \pi^2 \alpha')^{\frac{1 + p}{2}}} \int_0^\infty \frac{d t}{  t^{\frac{9 -p}{2}}} \frac{e^{- \frac{y^2}{2\pi\alpha' t}} \, \theta^4_2 \left(\left.\frac{ i \nu_0}{2}\right| it\right)} {\eta^9 (it) \, \theta_1 (i \nu_0 | it)},\nn
&=& \frac{2^4\, n_1 n_2 V_{p + 1} \sqrt{(1 - f^2 + g^2)(1 - f'^2 + g'^2)} \, \cosh^4 \frac{\pi \nu_0}{2}}{(8 \pi^2 \alpha')^{\frac{1 + p}{2}}} \int_0^\infty \frac{d t} { t^{\frac{9 - p}{2}}} e^{- \frac{y^2}{2\pi\alpha' t}} \nn
&\,& \times\prod_{n = 1}^\infty  \frac{\left[1 + 2 |z|^{2n} \cos \pi \nu_0 + |z|^{4n}\right]^4 }{(1 - |z|^{2n})^6 \left[1 - 2 |z|^{2n} \cosh 2\pi \nu_0 + |z|^{4n}\right]},
\eea 
 where in obtaining the first equality from the second equality of \eqref{eg-totalamplit} we have used the identities $\theta_1 (1 + \nu |\tau) = - \theta_1 (\nu | \tau) = \theta_1 (- \nu |\tau)$ and $\theta_1 (\frac{1 + \nu}{2} |\tau) = \theta_2 (\frac{\nu}{2} |\tau) $.  As in the previous cases, the large $y$ amplitude gives an attractive interaction for $p < 7$.  For small $y$, we need to pass this amplitude to the open string annulus one via the Jacobi transformation $t \to t' = 1/t$. Here in addition to the relations given in \eqref{jacobi} for the Dedekind $\eta$-function and the $\theta_1$-function, we need also the following for $\theta_2$-function as
 \be \label{theta2}
 \theta_2 (\nu|\tau) = \frac{1}{(- i \tau)^{1/2}} e^{- i \pi \nu^2/\tau} \theta_4 \left(\left.\frac{\nu}{\tau} \right|  -  \frac{1}{\tau}\right).
 \ee  
 The open string annulus amplitude is then 
 \bea\label{eg-nu02annulusamplit}
\Gamma &=&  \frac{2 \, n_1 n_2 V_{p + 1} \sqrt{ (f - f')^2 +  (g f' - f g')^2 - (g - g')^2 }}{(8 \pi^2 \alpha')^{\frac{1 + p}{2}}} \int_0^\infty \frac{d t'}{ t'^{\frac{1 + p}{2}}} \frac{e^{- \frac{y^2 t'}{2\pi\alpha' }} \, \theta^4_4 \left(\left. \frac{ \nu_0 t'}{2}\right| it' \right)}{\eta^9 (it') \, \theta_1 (\nu_0 t'| it')},\nn 
 &=& \frac{n_1 n_2 V_{p + 1} \sqrt{ (f - f')^2 +  (g f' - f g')^2 - (g - g')^2 }}{(8 \pi^2 \alpha')^{\frac{1 + p}{2}}} \int_0^\infty \frac{d t}{t^{\frac{1 + p}{2}}} e^{- \frac{y^2 t}{2\pi\alpha' }}\,  \frac{e^{\pi t}}{\sin\pi \nu_0 t}\nn
 &\,& \times \prod_{n = 1}^\infty \frac{\left[ 1 - 2 |z|^{2n - 1} \cos \pi \nu_0 t + |z|^{2 (2n - 1)}\right]^4}{(1 - |z|^{2n})^6 \left(1 - 2 |z|^{2n} \cos 2\pi \nu_0 t + |z|^{4n}\right)},
 \eea  
 where in the second equality we have dropped the prime on $t$ and again $|z| = e^{- \pi t} < 1$.  There are two dramatic differences from the previous subcase \eqref{egnu0-annulusamplit}: 1) The integrand has a factor $\sin\pi \nu_0 t$  in its denominator but without the presence of $\sin^4 \pi \nu_0 t/2$ in the numerator.  This $\sin\pi \nu_0 t$ factor gives then an infinite number of simple poles of the integrand along the positive $t$-axis at 
 \be\label{eg-nu02sp}
  t_k = \frac{k}{\nu_0}, \quad {\rm with}\,\,  k = 1, 2, \cdots.
  \ee
    2) There is an extra exponential factor $e^{\pi t}$ in the integrand which indicates an open string tachyon mode and the onset of tachyonic instability will occur when $y \le \pi \sqrt{2\alpha'}$.  These two features are precisely the ones for which one will see when each set of Dp branes carries an electric flux and a magnetic one but the two do not share a common field strength index. We will discuss this in the following section.
 
  Our previous examples already show that the electric flux(es) are responsible for the open string pair production while the magnetic one(s) are for the open string tachyon mode. Then the question is how to understand the appearance of the open string tachyon mode in the present subcase.  The simplest is to note that our original $\nu$-parameter is given as $\nu = 1 - i \nu_0$, which is complex. The real part $`1'$, which is due to magnetic fluxes, actually gives rise to the factor $e^{\pi t}$, therefore the open string tachyon mode.  Let us trace this.  Note that $1 - f f' + g g' < 0$ along with $(g - g')^2 < (f - f')^2 + (g f' - f g')^2$ gives $\nu = 1 - i\nu_0$. However, the real part $`1'$ is precisely due to $1 - f f' + g g' < 0$.  In the one-flux case considered in \cite{Lu:2009au}, we have either $f \neq 0, g = 0,  f' = 0, g' \neq 0$ or the other way around, then $1 - f f' + g g' = 1$ which can never be less than zero. Therefore this is consistent with what had been found there. That $1 - f f' + g g' < 0$ can hold is precisely due to the presence of the magnetic flux(es). If both $g = g' = 0$,  then $1 - f f' + g g' < 0$ would imply $1 < f f'$ which cannot be true since from $1 - f^2 > 0$ and $1 - f'^2 > 0$ we can have $f^2 f'^2 < 1$.  Let us now assume one of them being zero, say, $g' = 0$.  We then still need  $  f f' > 1$ from $1 - f f' + g g' < 0$.  From $1 - f^2 + g^2 > 0$ and $1 - f'^2 > 0$, we have $|f f'| < \sqrt{1 + g^2}$ which can be consistent with $f f' > 1$.  In this case,  all we need is to have $1 < f f' < \sqrt{1 + g^2}$.  If both $g$ and $g'$ are non-zero,  we have $f f' > 1 + g g'$ from $1 - f f' + g g' < 0$. From $1 - f^2 + g^2 > 0$ and $1 - f'^2 + g'^2 > 0$,  we have $|f f'| < \sqrt{(1 + g^2) (1 + g'^2)}$.  If $f f' < 0$, we then have $|f f'| < |1 + g g'|$ which can be consistent with $|f f'| < \sqrt{(1 + g^2) (1 + g'^2)}$ since $\sqrt{(1 + g^2) (1 + g'^2)} > |1 + g g'|$.  If $f f' > 0$,  then all we need is $1 + g g' < f f' < \sqrt{(1 + g^2) (1 + g'^2)}$.  In other words, so long there is a magnetic flux present, $1 - f f' + g g' < 0$ can hold, which gives rise to the real part $`1'$ in $\nu = 1 - i\nu_0$, therefore the open string tachyon mode. 
 
     As before, the simple poles \eqref{eg-nu02sp} give rise to the open string pair production at each of them  and the pair production rate can be calculated, by the same token, to be
 \bea\label{eg-nu02rate}
 {\cal W} &=& - \frac{2 {\rm Im} \Gamma}{V_{p + 1}}, \nn    
 &=& \frac{2 n_1 n_2 \left[(f - f')^2 + (f g' - g f')^2 - (g - g')^2\right]^{\frac{1}{2}}}{\nu_0 \, (8 \pi^2 \alpha')^{\frac{1 + p}{2}}} \sum_{k = 1}^\infty ( - )^{k - 1} \left(\frac{\nu_0}{k}\right)^{\frac{1 + p}{2}}
 e^{- \frac{k (y^2 - 2 \pi^2 \alpha')}{2 \pi \nu_0 \alpha'}}\nn
 &\,&\times  \prod_{n = 1}^\infty \left( \frac{1 - ( - )^k e^{ - \frac{(2n - 1) k \pi}{\nu_0}}}{1 - e^{- \frac{2 n k \pi}{\nu_0}}}\right)^8.
 \eea
 For large $\nu_0$, the discussion goes the same as before and  we will not repeat it here.  Our focus here is on $\nu_0 \ll 1$ and we will see the first example of rate enhancement discussed in the present paper.  When $\nu_0 \ll 1$, the rate can be approximated by the leading $k = 1$ term  as
 \be\label{eg-smalln02}
 {\cal W} \approx \frac{2 \pi n_1 n_2 \sqrt{(1 - f^2 + g^2)(1 - f'^2 + g'^2)}}{(8 \pi^2 \alpha')^{\frac{1 + p}{2}}} \nu_0^{\frac{1 + p}{2}} e^{- \frac{y^2 - 2\pi^2 \alpha'}{2 \pi \nu_0 \alpha'}}.
 \ee
 We now compare the present pair production rate with the one given  in \eqref{eg-ospr}. For this,  we assume the same $\nu_0$ in both cases and also the same $ \sqrt{(1 - f^2 + g^2)(1 - f'^2 + g'^2)}$ factor\footnote{\label{nu02case}There is no problem with this assumption. For checking this easily, we set $f = a \sinh\theta, g = a \cosh\theta$ and $f' = a' \sinh\theta', g' = a' \cosh\theta'$ from the conditions $1 - f^2 + g^2 > 0$ and $1 - f'^2 + g'^2 > 0$, respectively, for the former subcase.  For the present subcase, we use a bar on each of them  to make a distinction in notations.  Then the assumption gives two conditions: $a^2 + a'^2 + a^2 a'^2 = \bar a^2 + \bar a'^2 + \bar a^2 \bar a'^2$ and $2 = - a a' \cosh (\theta - \theta') - 
 \bar a \bar a' \cosh (\bar \theta - \bar \theta')$.  For the former case, $\cosh\pi\nu_0 > 1$ from \eqref{eg-nu0parameter} gives  $1 + a a' \cosh (\theta - \theta') > \sqrt{(1 + a^2)(1 + a'^2)}$ which implies $a a' > 0$. By the same token, the present subcase from \eqref{eg-nu02parameter} gives $ - \bar a \bar a' \cosh (\bar \theta - \bar\theta') > 1 + \sqrt{(1 + \bar a^2)(1 + \bar a'^2)} > 2$ which implies $\bar a \bar a' < 0$. If we set $\bar a = a, \bar a' = -  a'$, for example, the above first equation is satisfied. The second one can also be satisfied by choosing $\bar\theta - \bar\theta' = \cosh^{-1} (2/aa' + \cosh(\theta - \theta'))$ since $- \bar a \bar a' \cosh(\bar \theta - \bar\theta') = a a' \cosh(\bar \theta - \bar \theta') > 2$. }.  Then the present rate over the previous one gives a factor $e^{\pi/\nu_0}/8$, which can be very large for $\nu_0 \ll 1$, a great enhancement.   Note that the smallest $p = 2$ gives the largest rate when the fluxes are the same.  For separation $y = \pi \sqrt{2 \alpha'} + \Delta \sqrt{\alpha'}$ with $\Delta \ll \sqrt{2}\, \nu_0$,  the rate \eqref{eg-smalln02} is
 \be\label{eg-smallnu02atsy}
 (2 \pi \alpha')^{\frac{1 + p}{2}} {\cal W} \approx \frac{2 \pi n_1 n_2 \sqrt{(1 + a^2)(1 + a'^2)}} {(4\pi)^{\frac{1 + p}{2}}} \nu_0^{\frac{1 + p}{2}},
 \ee
 where we have expressed $f, g$ and $f', g'$ in terms of their respective $a, \theta$ and $a', \theta'$ as given in footnote \eqref{nu02case}. For small $\nu_0$, since both $|a|$ and $|a'|$ with $a a' < 0$ can still take large values, so this rate can still be large. 
 
 In summary, we have learned for various systems considered in this section that the electric flux gives rise to the open string pair production while the magnetic one gives rise to a tachyon mode, which can have the onset of tachyon instability when the brane separation is small, in terms of the open string annulus diagram description.  In the case of one electric flux and one magnetic one carrying by each set of Dp-branes considered in this section, when the two fluxes satisfy certain relations specified in the subcase 3) in subsection 3.2,  we find that the open string pair production can be significantly enhanced and this may have a potential realistic application which we will discuss in the discussion and conclusion section. 
 
 The rate enhancement found in subcase 3 is due to the sign change of $1 - f f' + g g'$ from positive to negative in comparison to the subcase 2 in subsection 3.2.  So we have the sign change of the R-R contribution to the amplitude and this appears to be like the brane/anti-brane system against brane/brane system in spirit. 
The following discussion indicates that this is not the case. This sign change of $1 - f f' + g g'$ is entirely due to the added electric and magnetic fluxes and the original two sets of Dp branes are not changed at all (just two sets of Dp branes, not one set of Dp and one set of anti-Dp).   As stressed in subcase 3, in addition to the electric flux  $f$  on one set of Dp and the electric flux $f'$ on the other set of Dp, we have to have at least one magnetic flux present on one of the two sets of Dp branes to have the enhancement to occur.  As discussed in subcase 3, we can set $g' = 0$ but keep $g \neq 0$. Given $ |f| < \sqrt{1 + g^{2}}$ and $|f'| < 1$, we can have  $1 < f f' < \sqrt{1 + g^{2}}$ for which the two electric fluxes point to the same direction and now $1 - f f' + g g' < 0$.   As mentioned in footnote \ref{fn2} of this paper, it is well-known that a constant worldvolume electric flux stands for the fundamental string while a constant magnetic flux stands for a $D(p - 2)$ brane inside the Dp brane. Given these, the Dp carrying the constant electric flux $f'$ stands for a 1/2 BPS non-threshold bound state (F, Dp) while the one carrying the constant electric flux $f$ and the magnetic flux $g$ stands for a 1/2 BPS non-threshold bound state ((F, D(p - 2)), Dp) as mentioned in footnote \ref{fn8}. The Dp in (F, Dp) is identical to the Dp in ((F, D(p -2)), Dp) and if we restrict $f f' > 0$, the F in (F, Dp) points to the same direction as the F in ((F, D(p -2)), Dp).  If we keep fixed  both $f'$ and $g$, we have the subcase 2 if $0 < f f' < 1$ for which $1 - f f' > 0$ and the subcase 3 if $1 < f f' < \sqrt{1 + g^{2}}$ for which $1 - f f' < 0$.  The two subcases differ only by the change of the magnitude of the electric flux $f$ and in either case the $((F, D(p - 2)), Dp)$ bound state is not the anti-system of (F, Dp) in the usual sense.  The sign change of R-R amplitude is due to the combined result of interactions of constituent branes in the two bound states. The present system has the advantage over the brane/anti-brane one in that it has a minor instability rather than highly unstable and as such the enhanced open string pair production can have the potential to be detected by an observer living on set of the branes.  

 We would like to stress that the open string pair production as well as its enhancement and the tachyon mode are due to the open strings connecting the two sets of the Dp branes, not the ones with their both ends on the same set of Dp branes.  Since each set of the Dp branes with fluxes by themselves are still 1/2 BPS,  we don't expect each set as an isolated system to have the open string pair production to occur even when the flux(es) they carry are electric.  This is also consistent with the fact that there is no pair production for neutral open string \cite{Bachas:1992bh,Porrati:1993qd}. In the following section, we will provide more evidence to support what has been found in this section when each set of Dp branes carry two fluxes with structures different from what has been considered in this section. 
 
 \section{The $8 \ge p \ge 3$ case}
 In this section, we will address the same issues as in the previous one but with the following flux structures,
 \be\label{eg3-structure}
\hat F =\left( \begin{array}{cccccc}
0 & - f & 0 &0 &0& \ldots\\
f&0&0&0&0&\ldots\\
0&0&0& -g&0&\ldots\\
0&0&g&0&0&\ldots\\
0&0&0&0&0&\ldots\\
 \vdots&\vdots&\vdots&\vdots&\vdots&\ddots
 \end{array}\right)_{(1 + p)\times (1 + p)},
 \ee
  or
  \be\label{gg3-structure}
   \hat F =\left( \begin{array}{cccccc}
0 & 0 & 0&0 & 0&\ldots\\
0&0&- g_1&0&0&\ldots\\
0&g_1&0&-g_2&0&\ldots\\
0&0&g_2&0&0&\ldots\\
0&0&0&0&0&\ldots\\
 \vdots&\vdots&\vdots&\vdots&\vdots&\ddots
 \end{array}\right)_{(1 + p)\times (1 + p)}.
  \ee 
So we will also have two subcases to consider. In the first subcase, we have the electric flux $\hat F_{01} = - \hat F_{10} = -f$ and the magnetic flux $\hat F_{23} = - \hat F_{32} = - g$ with the rest vanishing. The two non-vanishing fluxes do not share a common field strength index.  While for the second subcase, we have two magnetic fluxes: the magnetic flux 
$\hat F_{12} = - \hat F_{21} = - g_1$ and the other magnetic one $\hat F_{23} = - \hat F_{32} = - g_2$. These two magnetic fluxes share a common field strength index $`2'$. In what follows, we will consider each in order.   
 
 \subsection{The electric-magnetic case}
 We consider the first subcase with  two fluxes: one electric and the other magnetic.  The $p = 3$ case has already been studied in a recent paper \cite{Lu:2017tnm} by the present author. We here discuss the general $p$  for $3 \le p \le 8$. Specifically, we have one set of Dp branes carrying the flux $\hat F'$ and the other carrying the flux $\hat F$ as
 \be\label{eg3case}
\hat F' =\left( \begin{array}{cccccc}
0 & - f' & 0 &0 &0& \ldots\\
f'&0&0&0&0&\ldots\\
0&0&0& -g'&0&\ldots\\
0&0&g'&0&0&\ldots\\
0&0&0&0&0&\ldots\\
 \vdots&\vdots&\vdots&\vdots&\vdots& \end{array}\right),\quad     \hat F =\left( \begin{array}{cccccc}
0 & - f & 0 &0 &0& \ldots\\
f&0&0&0&0&\ldots\\
0&0&0& -g&0&\ldots\\
0&0&g&0&0&\ldots\\
0&0&0&0&0&\ldots\\
 \vdots&\vdots&\vdots&\vdots&\vdots&\ddots
 \end{array}\right),
  \ee
 where both of them are $(1 + p) \times (1+ p)$ matrices. With them, as before, the eigenvalues can be determined to be
 \bea\label{eq3-eigenv}
 \lambda + \lambda^{-1} &=& 2 \frac{(1 + f^2)(1 + f'^2) - 4 f f'}{(1 - f^2)(1 - f'^2)},\nn
 \lambda' + \lambda'^{-1} &=& 2 \frac{(1 - g^2)(1 - g'^2) + 4 g g'}{(1 + g^2) (1 + g'^2)},
 \eea
 where we have set $\lambda_0 = \lambda, \lambda_1 = \lambda^{-1}, \lambda_2 = \lambda', \lambda_3 = \lambda'^{-1}$ and $\lambda_4 = \cdots = \lambda_p = 1$.  The matrix element for zero-mode in the R-R sector can also be determined to be
 \be\label{eg3-mme}
{}_{\rm 0R}\langle B', \eta'| B, \eta\rangle_{\rm 0R}  = - \frac{ 2^4 \,(1 - f f' )(1 + g g')}{\sqrt{(1 - f^2)(1 - f'^2)(1 + g^2)(1 + g'^2) }} \delta_{\eta' \eta, +}.
\ee 
 With these, we  have, from \eqref{amplitudensns}, 
 \be\label{eg3-amplitudensns}
\Gamma_{\rm NSNS} (\eta'\eta) = \frac{n_1 n_2 V_{p + 1} \left[ (1 - f^2)(1 - f'^2)(1 + g^2)(1 + g'^2)\right]^{\frac{1}{2}}}{(8 \pi^2 \alpha')^{\frac{1 + p}{2}}} \int_0^\infty \frac{d t}{ t^{\frac{9 - p}{2}}} \frac{e^{- \frac{y^2}{2\pi\alpha' t}}}{|z|} \prod_{n = 1}^\infty {\cal A}_n (\eta\eta'),
\ee
 where we have defined 
 \bea\label{eg3-A}
 {\cal A}_n (\eta\eta') &=&  \left(\frac{1  + \eta'\eta |z|^{2n - 1}}{1 - |z|^{2n}}\right)^4  \frac{(1 + \eta'\eta\, \lambda  |z|^{2 n - 1})(1 + \eta'\eta\, \lambda^{-1}  |z|^{2 n - 1})}{(1 - \lambda |z|^{2n})(1 - \lambda^{-1} |z|^{2n})} \nn
 &\,&\times  \frac{(1 + \eta'\eta\, \lambda'  |z|^{2 n - 1})(1 + \eta'\eta\, \lambda'^{-1}  |z|^{2 n - 1})}{(1 - \lambda' |z|^{2n})(1 - \lambda'^{-1} |z|^{2n})}, 
 \eea
 in the NS-NS sector while in the R-R sector, we have, from \eqref{amplituderr},
 \be\label{eq3-amplituderr}
\Gamma_{\rm RR} (\eta'\eta) = - \frac{2^4 \,n_1 n_2 V_{p + 1}, (1 - f f')(1 + g g') }{(8\pi^2\alpha')^{\frac{1 + p}{2}}}  \delta_{\eta\eta', +} \int_0^\infty \frac{dt }{ t^{\frac{ 9 - 9}{2}}} \, e^{- \frac{y^2}{2\pi \alpha' t}}  \prod_{n = 1}^\infty {\cal B}_n ,
\ee 
 where we have used \eqref{eg3-mme} for the zero-mode matrix element and 
 \be\label{eg3-B} 
{\cal B}_n = \left(\frac{1 +  |z|^{2n}}{1 - |z|^{2n}}\right)^4  \frac{(1 +  \lambda |z|^{2n})(1 + \lambda^{-1} |z|^{2n})(1 +  \lambda' |z|^{2n})(1 + \lambda'^{-1} |z|^{2n})}{(1 - \lambda |z|^{2n})(1 - \lambda^{-1} |z|^{2n})(1 - \lambda' |z|^{2n})(1 - \lambda'^{-1} |z|^{2n})}.
\ee
So the GSO projected amplitude in the NS-NS sector is
\bea\label{eg3-amplitnsns}
\Gamma_{\rm NSNS} &=& \frac{1}{2} \left[\Gamma_{\rm NSNS} ( + ) - \Gamma_{\rm NSNS} (-) \right],\nn
&=& \frac{n_1 n_2 V_{p + 1} \left[(1 - f^2)(1 - f'^2)(1 + g^2)(1 + g'^2)\right]^{\frac{1}{2}}}{2 (8 \pi^2 \alpha')^{\frac{1 + p}{2}}} \int_0^\infty\frac{ d t}{t^{\frac{9 - p}{2}}}\frac{ e^{- \frac{y^2}{2\pi\alpha' t}}}{ |z|} \left[\prod_{n = 1}^\infty {\cal A}_n ( + )\right.\nn
&\,&\left. - \prod_{n = 1}^\infty {\cal A}_n (-)\right],
\eea
while the GSO projected amplitude in the R-R sector is
\bea\label{eg3-amplitrr}
\Gamma_{\rm RR} &=& \frac{1}{2} \left[\Gamma_{\rm RR} (+) + \Gamma_{\rm RR} (-)\right],\nn
&=& - \frac{2^3 \,n_1 n_2 V_{p + 1}, (1 - f f')(1 + g g') }{(8\pi^2\alpha')^{\frac{1 + p}{2}}}   \int_0^\infty \frac{dt}{ t^{\frac{ 9 - p}{2}}} \, e^{- \frac{y^2}{2\pi \alpha' t}}  \prod_{n = 1}^\infty {\cal B}_n.
\eea
We have then the total amplitude 
\bea\label{eg3-totalamplitude}
\Gamma &=& \Gamma_{\rm NSNS} + \Gamma_{\rm RR}\nn
&=& \frac{n_1 n_2 V_{p + 1} \left[(1 - f^2)(1 - f'^2)(1 + g^2)(1 + g'^2)\right]^{\frac{1}{2}}}{2 (8 \pi^2 \alpha')^{\frac{1 + p}{2}}} \int_0^\infty \frac{d t}{t^{\frac{9 - p}{2}}} e^{- \frac{y^2}{2\pi\alpha' t}} \left[|z|^{-1}\left(\prod_{n = 1}^\infty {\cal A}_n ( + )\right.\right.\nn
&\,& \left. \left.- \prod_{n = 1}^\infty {\cal A}_n (-)\right) -  \frac{2^4 \, (1 - f f')(1 + g g')}{\sqrt{ (1 - f^2)(1 - f'^2)(1 + g^2)(1 + g'^2)}}  \prod_{n = 1}^\infty {\cal B}_n\right],\nn
\eea
where ${\cal A}_n (\pm)$ and ${\cal B}_n$ are defined in \eqref{eg3-A} and \eqref{eg3-B}, respectively.  Let us try to express this amplitude in terms of various $\theta$-functions and the Dedekind $\eta$-function. For this, let us define,
\be\label{nu-nu'-parameter}
\lambda = e^{2\pi i \nu},  \quad \lambda' = e^{2\pi i \nu'}.
\ee
Using \eqref{eq3-eigenv}, we have
\bea\label{eg3-nu-nu'}
\cosh\pi \nu_0 &=& \frac{1 - f f'}{\sqrt{(1 - f^2)(1 - f'^2)}}, \quad \sinh\pi\nu_0 = \frac{|f - f'|}{\sqrt{(1 - f^2) (1 - f'^2)}},\nn
\cos\pi \nu'_0 &=& \frac{1 + g g'}{\sqrt{(1 + g^2)(1 + g'^2)}}, \quad \sin\pi\nu'_0 = \frac{|g - g'|}{\sqrt{(1 + g^2)(1 + g'^2)}},
\eea
where we have defined $\nu = i \nu_0$ with $0 < \nu_0 < \infty$ and $\nu' = \nu'_0$ with $0 < \nu'_0 < 1$. Note that when either $f$ or $f'$ reaches its critical value\footnote{When both reach their critical values, we can set $f = \pm 1 \mp \epsilon, f' = \pm 1 \mp \epsilon'$ with $\epsilon \to 0$ and $\epsilon' \to 0$ and $\nu_0 \to \infty$ if $\epsilon/\epsilon' \to 0\, {\rm or} \, \infty$.}  of unity, $\nu_0 \to \infty$.  With these, the total amplitude \eqref{eg3-totalamplitude} can be expressed as
\bea\label{eg3-totalamplit}
\Gamma &=& \frac{2 \,i \,n_1 n_2 V_{p + 1} |f - f'||g - g'|}{ (8 \pi^2 \alpha')^{\frac{1 + p}{2}}} \int_0^\infty \frac{d t}{ t^{\frac{9 - p}{2}}} e^{- \frac{y^2}{2\pi\alpha' t}}\nn
&\,& \times \frac{\theta_3^2 (0 | it) \theta_3 (\nu|it)\theta_3 (\nu'|it) - \theta_4^2 (0 |it) \theta_4 (\nu |it) \theta_4 (\nu' | it) - \theta^2_2 (0 | it) \theta_2 (\nu |it) \theta_2 (\nu'|it)}{\eta^6 (it) \theta_1 (\nu|it) \theta_1 (\nu' | it)},\nn
&=&  \frac{2^2 \,i \,n_1 n_2 V_{p + 1} |f - f'||g - g'|}{ (8 \pi^2 \alpha')^{\frac{1 + p}{2}}} \int_0^\infty \frac{d t} {t^{\frac{9 -p}{2}}} e^{- \frac{y^2}{2\pi\alpha' t}}\frac{\theta^2_1 \left(\left.\frac{\nu -\nu'}{2}\right|  it \right) \theta_1^2 \left(\left.\frac{\nu +\nu'}{2}\right| it\right)}{\eta^6(it) \theta_1 (\nu|it) \theta_1 (\nu'|it)},
\eea
where in the last equality we have used the following identity 
\bea\label{eg3-jacobi}
2 \theta_1^2 \left(\left.\frac{\nu - \nu'}{2}\right|\tau\right)  \theta_1^2 \left(\left.\frac{\nu + \nu'}{2}\right|\tau\right) &=& \theta^2_3 (0|\tau) \theta_3 (\nu|\tau) \theta_3 (\nu' |\tau) - \theta^2_4 (0|\tau) \theta_4 (\nu|\tau) \theta_4(\nu'|\tau) \nn
&\,& - \theta^2_2 (0|\tau) \theta^2_2 (\nu|\tau)\theta^2_2 (\nu'|\tau),
\eea 
which is again a special case of more general identity given in \cite{whittaker-watson}. We can now set $\nu = i\nu_0$ and $\nu' = \nu'_0$ in the last equality of \eqref{eg3-totalamplit} and use the explicitt expressions for the $\theta_1$ function and the Dedekind $\eta$ function to have
\bea\label{eg3-totalamplitexplicit}
\Gamma &=&  \frac{2^2 n_1 n_2 V_{p + 1} [ (1 - f^2)(1 - f'^2)(1 + g^2)(1 + g'^2)]^{\frac{1}{2}} }{ (8 \pi^2 \alpha')^{\frac{1 + p}{2}} (\cosh\pi\nu_0 - \cos\pi\nu'_0)^{-2}} \int_0^\infty \frac{d t}  {t^{\frac{9 - p}{2}}} e^{- \frac{y^2}{2\pi\alpha' t}} \nn
& &\times \prod_{n = 1}^\infty \frac{\left[(1 - 2 |z|^{2n} e^{- \pi \nu_0} \cos\pi\nu'_0 + e^{- 2 \pi \nu_0} |z|^{4n})(1 - 2 |z|^{2n} e^{\pi \nu_0} \cos\pi\nu'_0 + e^{2\pi \nu_0} |z|^{4n})\right]^2}{(1 - |z|^{2n})^4 \left[1 - 2 |z|^{2n} \cosh2\pi \nu_0 + |z|^{4n}\right] \left[1 - 2 |z|^{2n} \cos\pi\nu'_0 + |z|^{4n}\right]}.\nn
\eea
For large brane separation $y$, this amplitude gives a finite positive one $\Gamma \propto 1/y^{7 - p} > 0$ for $p < 7$, implying an attractive interaction, as expected. As can be seen, $\Gamma = 0$ only if $\cosh\pi \nu_0 - \cos\pi \nu' = 0$ whose only solution is $\nu_0 = \nu'_0 = 0$. This gives $f = f', g = g'$ and the underlying system is still a 1/2 BPS state.

For small brane separation $y$, we expect also the open string pair production to occur since we have an electric flux present and the best description is in terms of the open string annulus variable which can be obtained via the Jacobi transformation $t \to t' = 1/t$.  Using the relations \eqref{jacobi} for the $\theta_1$-function and the Dedekind $\eta$-function, 
we have the open string annulus amplitude from the second equality in \eqref{eg3-totalamplit} as
\bea\label{eg3-totalannulusamplit}
\Gamma  &=&  \frac{2^2 \,i \,n_1 n_2 V_{p + 1} |f - f'||g - g'|}{ (8 \pi^2 \alpha')^{\frac{1 + p}{2}}} \int_0^\infty \frac{d t'} {t'^{\frac{p -1}{2}}} e^{- \frac{y^2 t'}{2\pi\alpha' }}\frac{\theta^2_1 \left(\left.\frac{\nu_0 + i\nu'_0}{2} t'\right|  it' \right) \theta_1^2 \left(\left.\frac{\nu_0 - i \nu'_0}{2} t'\right| it'\right)}{\eta^6(it') \theta_1 (\nu_0 t'|it') \theta_1 (i\nu'_0 t'|it')},\nn
&=& \frac{2^2 \,n_1 n_2 V_{p + 1} |f - f'||g - g'|}{ (8 \pi^2 \alpha')^{\frac{1 + p}{2}}} \int_0^\infty \frac{d t} {t^{\frac{p -1}{2}}} e^{- \frac{y^2 t}{2\pi\alpha' }} \frac{(\cosh \pi \nu'_0 t - \cos\pi \nu_0 t)^2}{\sin\pi\nu_0 t \sinh\pi\nu'_0 t}\nn
&\times& \prod_{n = 1}^\infty \frac{\left[(1 - 2 |z|^{2n} e^{ - \pi \nu'_0 t} \cos\pi\nu_0 t + |z|^{4n} e^{- 2 \pi \nu'_0 t})(1 - 2 |z|^{2n} e^{\pi \nu'_0 t} \cos\pi \nu_0 t + |z|^{4n} e^{2 \nu'_0 t})\right]^2}{(1 - |z|^{2n})^6 \left(1 - 2 |z|^{2n} \cos2\pi \nu_0 t + |z|^{4n}\right) \left(1 - 2 |z|^{2n} \cosh2\pi \nu'_0 t + |z|^{4n}\right)}, \nn
\eea
where we have set $\nu = i \nu_0, \nu' = \nu'_0$ and in the second equality we have dropped the prime on $t$ and $|z| = e^{- \pi t} < 1$.  Let us examine the behavior of the integrand in the second equality above.  The integrand has the following divergent behavior 
\be\label{blowup}
\lim_{t \to \infty} e^{- \frac{y^2 t}{2\pi\alpha' }} \frac{(\cosh \pi \nu'_0 t - \cos\pi \nu_0 t)^2}{ \sinh\pi\nu'_0 t} \sim \lim_{t\to \infty}  e^{- \frac{(y^2 - 2 \pi^2 \nu'_0 \alpha') t}{2\pi \alpha'}} \to \infty,
\ee
if $y < \pi \sqrt{2\nu'_0 \alpha'}$, signaling the onset of tachyonic instability \cite{Sen:1999xm, Pesando:1999hm}.   The appearance of the exponential growing factor $e^{\pi \nu'_0 t}$ for large $t$ in the integrand indicates the existence of an open string tachyon mode which is due to the magnetic fluxes. This integrand blows up also when the factor  $\sin\pi \nu_0 t$ in the denominator vanishes along the positive $t$-axis at 
\be\label{eg3-sp}
t_k = \frac{k}{\nu_0}, \quad k = 1, 2, \cdots.
\ee
This blowing-up behavior actually gives rise to new physics at each of the infinite number of simple poles, indicating the production of an open string pair under the action of electric fluxes applied. This implies that the amplitude has an imaginary part. The pair production rate per unit Dp-brane worldvolume is the imaginary
part of the amplitude, which can be obtained as the sum of the residues of the poles of the integrand in \eqref{eg3-totalannulusamplit} times $\pi$ following \cite{Bachas:1992bh,Bachas:1995kx} and is given as
\bea\label{eg3-pprate}
 {\cal W} &=& - \frac{2 \,{\rm Im} \Gamma}{V_{p + 1}},\nn
&=& \frac{8\, n_1 n_2  |f - f'||g - g'|}{(8\pi^2 \alpha')^{\frac{1 + p}{2}}} \sum_{k = 1}^\infty (-)^{k - 1} \left(\frac{\nu_0}{k}\right)^{\frac{p - 3}{2}} \frac{\left[\cosh\frac{\pi k \nu'_0}{\nu_0} - (-)^k\right]^2}{k \,\sinh \frac{\pi k \nu'_0}{\nu_0}} \, e^{- \frac{ k\, y^2}{2\pi \alpha' \nu_0}}\nn
&\,&\times  \prod_{n = 1}^\infty  \frac{\left[1 -  (-)^k \, e^{- \frac{2 n k \pi}{\nu_0} (1 - \frac{\nu'_0}{2 n})}\right]^4 \left[1 - (-)^k  \, e^{- \frac{2 n k \pi}{\nu_0}(1 + \frac{\nu'_0}{2 n})}\right]^4}{\left(1 - e^{- \frac{2 n k \pi}{\nu_0}}\right)^6 \left[1 -   \, e^{- \frac{2 n k \pi}{\nu_0} (1 - \nu'_0/ n)}\right] \left[1 -  \, e^{- \frac{2 n k \pi}{\nu_0}(1 + \nu'_0 / n)}\right]}.
\eea  
We now discuss certain properties of this rate.  First the odd $k$ gives a positive contribution to the rate while the even $k$ gives a negative one. The $k = 1$ gives the leading positive contribution to the rate. For given fluxes (therefore also $\nu_0$ and $\nu'_0$),  the larger the brane separation $y$ is,  the larger the mass of the created open string (the string tension times the brane separation) is.  Moreover  for given $y$, the larger the $k$ is, the larger the string tension (k times the fundamental string tension) is and so also the larger the mass is.  Either case implies more difficulty to produce the open string pair.  This is reflected by the exponentially suppressed factor  ${\rm exp} [- k \, y^2/(2\pi \alpha' \nu_0)]$ in the rate \eqref{eg3-pprate}. Note that this rate appears valid only for $y > \pi \sqrt{2 \nu'_0 \alpha'}$ since the term for large $k$ would diverge, due to the open string tachyon mode mentioned in \eqref{blowup}.  Now for fixed $k$ and $y$, the parameters $\nu_0$ and $\nu'_0$ can be re-expressed from  \eqref{eg3-nu-nu'} as
\be\label{eg3-nu0-nu'0}
 \tanh\pi\nu_0 = \frac{|f - f'|}{1- f f'},\qquad  \tan \pi\nu'_0 = \frac{|g - g'|}{1 + g g'}.
\ee
Note that $0 < \nu_0 < \infty$ and $|f|, |f'| < 1$.  From the above, we can see that the larger $|f|$ and $|f'|$ with  $f \neq f'$ are, the larger $\nu_0$ is. This is particularly true if $f f' < 0$.   Moreover, when either $|f|$ or $|f'|$ reaches its critical value of unity, $\nu_0 \to \infty$.  When both reach their critical values but with $f f' < 0$, $\nu_0 \to \infty$.  When both reach their critical values but with $f f' > 0$, we can set $f = \pm 1 \mp \epsilon$ and $f' = \pm 1 \mp \epsilon'$ with both $ \epsilon \to 0^+, \epsilon' \to 0^+$.  For this case, $\nu_0 \to \infty$ only if $\epsilon/\epsilon' \to 0 \, {\rm or} \, \infty$.    For magnetic fluxes, we have $0 < \nu'_0 < 1$ and $ |g|, |g'| < \infty$. When both $|g|$ and $|g'|$ are very small or very large with $g g' > 0$,  $\nu'_0 \to 0$. When both are very large but with $g g' < 0$, $\nu'_0 \to 1$. So we have $0 < \nu'_0 < 1/2$ for $-1 < g g' < \infty$, $\nu'_0 = 1/2$ for $ g g' = - 1$ and $1/2 < \nu'_0 < 1$ for $ - \infty < g g' < - 1$.  The rate will be larger if we have a larger $\nu'_0$ and a larger $|g - g'|$.  When $g g' < 0$, the larger both $|g|$ and $|g'|$ are, the larger $\nu'_0 $  ($1/2 < \nu'_0 < 1$) and $|g - g'|$ are.  With respect to the above fluxes, we have three cases to consider, 
\be\label{threecase}
{\rm Case  \,\, I:} \quad \frac{\nu'_0}{\nu_0} \ll 1, \quad {\rm Case\,\, II:} \quad \frac{\nu'_0}{\nu_0} \sim {\cal O} (1), \quad {\rm Case\,\, III:}\quad  \frac{\nu'_0}{\nu_0} \gg 1.
\ee    
We here discuss each of them in order.\\
\noindent
{\bf Case I:}  Unless both $|g|$ and $|g'|$ are\footnote{When both $|g|$ and $|g'|$ are very small or very large but with $g g' > 0$, Case I needs only a finite $\nu_0$ to hold.  We will not discuss this situation since it will not give an interesting and useful rate unless $|g - g'|$ is very large for the latter.} very small or very large but with $g g' > 0$, we have in general $\nu'_0 \sim {\cal O} (1)$. This case requires a large $\nu_0$, therefore large electric fluxes $|f|$ and $|f'|$ with $f \neq f'$. From \eqref{eg3-pprate}, it is clear that the larger the $\nu_0$ is, the larger each term in the sum and so the larger the rate is.  Here each odd $k$ term gives a larger positive contribution while each even $k$ term gives also a larger but almost vanishing contribution to the rate (note that the even $k$ term is in general negative). This is expected. In particular, for any of the cases discussed above with the critical electric flux or fluxes and $\nu_0 \to \infty$, the rate blows up, giving  the onset of pair production instability.\\

\noindent 
{\bf  Case II:}  This case says $\nu_0 \sim {\cal O} (\nu'_0) \sim {\cal O} (1)$ and the only possible enhancement of the rate is due to the factor $|g - g'|$.  For a small $|g - g'|$, the rate is small too. For a large $|g - g'|$, the rate can be significant for a small brane separation but is still  small for a large brane separation.\\

\noindent 
{\bf Case III:}  This case must imply that $\nu_0 \ll 1$ since\footnote{We here assume that $\nu'_0$ is fixed in the range of $0 < \nu'_0 < 1$, not considering the case of $\nu'_0 \to 0$.} $0 < \nu'_0 < 1$. One in general would expect a vanishing small rate as the pure electric flux case given in \cite{Lu:2009yx} by the present author and his collaborators. It turns out that the story here is quite different and the added magnetic fluxes give an exponential enhancement of the rate via the tachyon mode discussed earlier.  A special $p = 3$ case has been reported recently by the present author in \cite{Lu:2017tnm}.  The simplified one-flux case was also given a while ago by the present author and his collaborator in \cite{Lu:2009au}.  We here give a discussion for a general $p$ and with fixed two fluxes, one electric and one magnetic, on each set of Dp branes, with the given requirement. In other words, 
we have here fixed $\nu'_0 \neq 0$ and $\nu_0 \neq 0$ with $\nu'_0/\nu_0 \gg 1$.  With a very small $\nu_0$, the infinite product for each $k$ in the sum in \eqref{eg3-pprate} can be approximated as unity. Moreover with $\nu'_0/\nu_0 \gg 1$, we can approximate the rate \eqref{eg3-pprate} as
\be\label{eg3-approx-pprate}
 {\cal W}  (\nu'_0 \neq 0) = \frac{8\, n_1 n_2  |f - f'||g - g'|}{(8\pi^2 \alpha')^{\frac{1 + p}{2}}} \sum_{k = 1}^\infty (-)^{k - 1} \frac{1}{k} \left(\frac{\nu_0}{k}\right)^{\frac{p - 3}{2}}  e^{- \frac{ k\, y^2}{2\pi \alpha' \nu_0}}\, e^{\frac{k \pi  \nu'_0}{\nu_0}},
\ee      
where the exponentially large factor ${\rm exp} ( k \pi \nu'_0 /\nu_0)$ is due to the open string tachyon mode discussed in \eqref{blowup}. Let us compare this rate,  for the same small $\nu_0$,  with the one without the presence of magnetic fluxes (i.e. $g,  g' = 0$ and $\nu'_0 = 0$) as given in 
\cite{Lu:2009yx}\footnote{This  can also be obtained from \eqref{eg3-pprate} by setting $g, g', \nu'_0 \to 0$.} as
\be\label{ee3-pprate}
{\cal W} (\nu'_0 = 0) \approx  \frac{32\, n_1 n_2  |f - f'| \,\nu_0 }{(8\pi^2 \alpha')^{\frac{1 + p}{2}} }\sum_{l = 1}^\infty  \left(\frac{\nu_0}{2 l - 1}\right)^{\frac{p - 3}{2}} \frac{1}{(2l -1)^2}\, e^{- \frac{ (2l -1) y^2}{2\pi \alpha' \nu_0} },
\ee
where we have set $k = 2 l - 1$ and the even $k$ doesn't contribute to this rate.  So it is clear for each odd $k = 2l -1$, there is a greatly enhanced factor
\be
\frac{{\cal W} ^l (\nu'_0 \neq 0)} {{\cal W}^l (\nu'_0 = 0)} = \frac{(2l - 1) |g - g'| e^{(2 l - 1) \pi \nu'_0/\nu_0}}{4 \nu_0},
\ee
where the superscript $`l$' denotes the l-th term in the corresponding rate summation.   For small enough $\nu_0$ and reasonable large $\nu'_0$, this enhancement can be very significant.  Now the corresponding rate can be approximated by the leading $k = 1$ or $l = 1$ term and the enhancement  is  
\be
\frac{{\cal W}  (\nu'_0 \neq 0)} {{\cal W} (\nu'_0 = 0)} = \frac{ |g - g'| e^{\pi \nu'_0/\nu_0}}{4 \nu_0}.
\ee
 Let us make the same sample numerical estimation of this enhancement as in \cite{Lu:2017tnm} for $p = 3$ to demonstrate its significance.  It has a value of $3.2\times 10^{35}$, a very significant enhancement, for $\nu_0 = 0.02$, $\nu'_0 = 0.5$. This can be achieved using \eqref{eg3-nu-nu'} via a moderate choice of $g_1 = - g_2 = 1$ (noting $|g_a| < \infty$) and $f_1 = 0.2$ with $f_2 = f_1 - \epsilon$ and $|f_1 - f_2| = |\epsilon| \approx \pi \nu_0 (1 - f_1^2) = 0.06 \ll 1$.  In spite of this, in order to be physically significant, the rate itself in string units needs to be large enough, not merely the enhancement  factor.  The rate in string units for the above sample case can be estimated to be 
\bea\label{eg3-rate-estimation} 
(2\pi \alpha')^{\frac{1+ p}{2}}  {\cal W} (\nu'_0 = 0.5) &\approx& \frac{n_1 n_2 |f - f'| |g - g'|}{2 \pi^2}\, \left(\frac{\nu_0}{4 \pi}\right)^{\frac{p - 3}{2}}\, e^{- \frac{y^2 - 2 \pi^2 \alpha' \nu'_0}{2\pi \alpha' \nu_0}} \nn
&\approx&  0.61\, (0.04)^{p - 3}\, e^{- \frac{y^2 -  \pi^2 \alpha' }{0.04 \pi \alpha' }},
\eea
with a typical choice of $n_1 = n_2 = 10$.   As discussed in \cite{Lu:2017tnm}, the rate for $p = 3$ is the largest and the rate for $p > 3$ is at least smaller by a factor of $(\nu_0/4 \pi)^{1/2} \approx 0.04$, i.e. two orders of magnitude smaller, for the sample case considered.   For $p = 3$, this rate $(2\pi \alpha')^2 {\cal W} (\nu'_0 = 0.5) = 0.61$, quite significant, at $y = \pi \sqrt{\alpha'} + 0^+ \approx \pi \sqrt{\alpha'}$ , a few times of string scale and before the onset of tachyon condensation, but decreases exponentially with the separation square $y^2$ for $y > \pi \sqrt{\alpha'}$.  For example, the rate becomes half of its maximal value at $y - \pi \sqrt{\alpha'} \approx  0.01 \sqrt{\alpha'}$, just $1\%$ of the string scale. 
We will come back to discuss the significance, implications and potential applications of this enhanced pair production rate later in section 6.  For now, we move to the second subcase in this section.

\subsection{The magnetic-magnetic case}
We consider the second subcase as given in \eqref{gg3-structure} with two fluxes, both magnetic, for $3 \le p \le 8$. Specifically,  we have one set of Dp branes carrying the flux $\hat F'$ and the other set carrying the flux $\hat F$ as
 \be\label{gg3case}
   \hat F' =\left( \begin{array}{cccccc}
0 & 0 & 0&0 & 0&\ldots\\
0&0&- g'_1&0&0&\ldots\\
0&g'_1&0&-g'_2&0&\ldots\\
0&0&g'_2&0&0&\ldots\\
0&0&0&0&0&\ldots\\
 \vdots&\vdots&\vdots&\vdots&\vdots&\ddots
 \end{array}\right), \quad  \hat F =\left( \begin{array}{cccccc}
0 & 0 & 0&0 & 0&\ldots\\
0&0&- g_1&0&0&\ldots\\
0&g_1&0&-g_2&0&\ldots\\
0&0&g_2&0&0&\ldots\\
0&0&0&0&0&\ldots\\
 \vdots&\vdots&\vdots&\vdots&\vdots&\ddots
 \end{array}\right), 
  \ee 
where both matrices are $(1 + p)\times (1 + p)$. With them, following the same steps as before, we have the eigenvalues 
\bea\label{gg3-eigen}
\lambda_1 \lambda_2 \lambda_3 &=& 1,\nn
\lambda_1 + \lambda_2 + \lambda_3 &=& \frac{1}{\lambda_1} + \frac{1}{\lambda_2} + \frac{1}{\lambda_3} = \lambda_2 \lambda_3 + \lambda_1 \lambda_3 + \lambda_1 \lambda_2,\nn
&=& \frac{3 (1 + g'_1 g_1 + g'_2 g_2)^2 - (g'_1 - g_1)^2 - (g'_2 - g_2)^2 - (g'_1 g_2 - g'_2 g_1)^2}{(1 + g'^2_1 + g'^2_2) (1 + g_1^2 + g_2^2)},
\eea
and $\lambda_0 = \lambda_4 = \cdots = \lambda_p = 1$. The zero-mode matrix element \eqref{0mme} for the present case  in the R-R sector can be determined to be
 \be\label{gg3-mme}
{}_{\rm 0R}\langle B', \eta'| B, \eta\rangle_{\rm 0R}  = - \frac{ 2^4 \,(1 + g_1 g'_1 + g_2 g'_2)}{\sqrt{(1 + g_1^2 + g_2^2) (1 + g'^2_1 + g'^2_2) }} \delta_{\eta' \eta, +}.
\ee 
We  have then the amplitude, from \eqref{amplitudensns}, in the NS-NS sector as
\bea\label{gg3-amplitudensns}
\Gamma_{\rm NSNS} (\eta \eta') &=& \frac{n_1 n_2 V_{p + 1} \left[(1 + g^2_1 + g^2_2)(1 + g'^2_1 + g'^2_2)\right]^{\frac{1}{2}}}{(8\pi^2 \alpha')^{\frac{1 + p}{2}}} \int_0^\infty \frac{d t}{t^{\frac{9 - p}{2}}} e^{ - \frac{y^2}{2\pi \alpha' t}} |z|^{ -1}\nn
&\,& \times \prod_{n = 1}^\infty \left(\frac{1 + \eta \eta' |z|^{2n - 1}}{1 - |z|^{2n}}\right)^5 \prod_{\alpha = 1}^3 \frac{1 + \lambda_\alpha \eta \eta'  |z|^{2n - 1}}{1 - \lambda_\alpha |z|^{2n}},
\eea
where each term involving the eigenvalues $\lambda_1, \lambda_2$ and $\lambda_3$ in the infinite product  can be simplified, using the relations given in \eqref{gg3-eigen}, as 
\be
\prod_{\alpha = 1}^3 \frac{1 + \lambda_\alpha \eta \eta'  |z|^{2n - 1}}{1 - \lambda_\alpha |z|^{2n}} = \frac{1 + \eta \eta' |z|^{2n - 1}}{1 - |z|^{2n}} \frac{(1 + \lambda\, \eta\eta' |z|^{2n - 1})(1 + \lambda^{-1}\, \eta \eta' |z|^{2n - 1})}{(1 - \lambda |z|^{2n}) (1 - \lambda^{-1} |z|^{2n})},
\ee
where 
\bea\label{gg3-lambda}
\lambda + \lambda^{-1} &=& \lambda_1 + \lambda_2 + \lambda_3 - 1,\nn
&=& 2  \frac{ (1 + g'_1 g_1 + g'_2 g_2)^2 - (g'_1 - g_1)^2 - (g'_2 - g_2)^2 - (g'_1 g_2 - g'_2 g_1)^2}{(1 + g'^2_1 + g'^2_2) (1 + g_1^2 + g_2^2)}.
\eea
The above NS-NS amplitude can now be expressed as
\be\label{gg3-amplitnsns}
\Gamma_{\rm NSNS} (\eta \eta') = \frac{n_1 n_2 V_{p + 1} \left[(1 + g^2_1 + g^2_2)(1 + g'^2_1 + g'^2_2)\right]^{\frac{1}{2}}}{(8\pi^2 \alpha')^{\frac{1 + p}{2}}} \int_0^\infty \frac{d t}{t^{\frac{9 - p}{2}}} \frac{e^{ - \frac{y^2}{2\pi \alpha' t}}}{ |z|} \prod_{n = 1}^\infty  A_n (\eta\eta'), 
\ee
where $A_n (\eta \eta')$ is defined in \eqref{A} but  for now with the $\lambda$ given by \eqref{gg3-lambda}. 
We have then the GSO-projected NS-NS amplitude 
\bea\label{gg3-gsoamplitnsns}
\Gamma_{\rm NSNS} &=& \frac{1}{2} \left[\Gamma_{\rm NSNS} (+) - \Gamma_{\rm NSNS} (-)\right],\nn
&=& \frac{n_1 n_2 V_{p + 1} \left[(1 + g^2_1 + g^2_2)(1 + g'^2_1 + g'^2_2)\right]^{\frac{1}{2}}}{2 (8\pi^2 \alpha')^{\frac{1 + p}{2}}} \int_0^\infty \frac{d t}{t^{\frac{9 - p}{2}}} \frac{e^{ - \frac{y^2}{2\pi \alpha' t}}} {|z|}\left[\prod_{n = 1}^\infty   A_n (+)\right.\nn
&\,& \left. - \prod_{n = 1}^\infty  A_n (-)\right].
\eea
By the same token, using \eqref{gg3-mme}, we have the amplitude, from \eqref{amplituderr}, in the R-R sector as
\be\label{gg3-amplitrr}
\Gamma_{\rm RR} (\eta \eta') = -  \frac{2^4\,n_1 n_2 V_{p + 1} (1 + g_1 g'_1 + g_2 g'_2)}{(8\pi^2 \alpha')^{\frac{1 + p}{2}}} \delta_{\eta\eta', +}  \int_0^\infty \frac{d t}{t^{\frac{9 - p}{2}}}\frac{ e^{ - \frac{y^2}{2\pi \alpha' t}}}{ |z|} \prod_{n = 1}^\infty B_n ,
\ee
where $B_n$ is defined in \eqref{B} but again for now with the $\lambda$ given  in \eqref{gg3-lambda}.
Then the GSO-projected amplitude in the R-R sector is
\bea\label{gg3-amplitrr}
\Gamma_{\rm RR} &=& \frac{1}{2} \left[\Gamma_{\rm RR} (+) + \Gamma_{\rm RR} (-)\right],\nn
 &=&-  \frac{2^3\,n_1 n_2 V_{p + 1} (1 + g_1 g'_1 + g_2 g'_2)}{(8\pi^2 \alpha')^{\frac{1 + p}{2}}} \int_0^\infty \frac{d t}{t^{\frac{9 - p}{2}}} \frac{e^{ - \frac{y^2}{2\pi \alpha' t}}}{ |z|}
 \prod_{n = 1}^\infty B_n.
\eea
We have then the total amplitude
\bea\label{gg3-totalamplitude}
\Gamma &=& \Gamma_{\rm NSNS} + \Gamma_{\rm RR},\nn
&=& \frac{n_1 n_2 V_{p + 1} \left[(1 + g^2_1 + g^2_2)(1 + g'^2_1 + g'^2_2)\right]^{\frac{1}{2}}}{2 (8\pi^2 \alpha')^{\frac{1 + p}{2}}} \int_0^\infty \frac{d t}{t^{\frac{9 - p}{2}}} e^{ - \frac{y^2}{2\pi \alpha' t}} \left[|z|^{-1} \left(\prod_{n = 1}^\infty A_n (+)\right.\right.\nn
&\,& \left.\left. - \prod_{n = 1}^\infty A_n (-)\right) - \frac{ 2^4 \,(1 + g_1 g'_1 + g_2 g'_2)}{\sqrt{(1 + g_1^2 + g_2^2) (1 + g'^2_1 + g'^2_2) }} \prod_{n = 1}^\infty B_n\right].
\eea
 As before, we define $\lambda =  e^{2 \pi i \nu}$ for the purpose of expressing this amplitude in a useful form which can facilitate its analysis. We then have from \eqref{gg3-lambda} 
\bea\label{gg3-nuparameter}
\cos\pi \nu'_0 &=& \frac{1 + g_1 g'_1 + g_2 g'_2}{\sqrt{(1 + g^2_1 + g^2_2)(1 + g'^2_1 + g'^2_2)}}, \nn
\sin\pi \nu'_0 &=& \sqrt{\frac{(g_1 - g'_1)^2 + (g_2 - g'_2)^2 + (g_1 g'_2 - g_2 g'_1)^2} {(1 + g^2_1 + g^2_2)(1 + g'^2_1 + g'^2_2)}},
\eea
where we have set $\nu = \nu'_0$ with $0 < \nu'_0 < 1$, denoting its magnetic nature as before. With this, the amplitude \eqref{gg3-totalamplitude} can now be expressed in terms of various $\theta$-functions and the Dedekind $\eta$-function as
\bea\label{gg3-totalamplit}
\Gamma &=& \frac{n_1 n_2 V_{p + 1} \left[(g_1 - g'_1)^2 + (g_2 - g'_2)^2 + (g_1 g'_2 - g_2 g'_1)^2 \right]^{\frac{1}{2}}}{ (8\pi^2 \alpha')^{\frac{1 + p}{2}}} \int_0^\infty \frac{d t}{t^{\frac{9 - p}{2}}} e^{ - \frac{y^2}{2\pi \alpha' t}}\nn
&&\times \frac{\theta^3_3 (0 | it) \theta_3 (\nu'_0 | it) - \theta^3_4 (0 | it) \theta_4 (\nu'_0 | it) - \theta^3_2 (0 | it) \theta_2 (\nu'_0 | it)}{\eta^9 (it) \theta_1 (\nu'_0 | it)},\nn
&=& \frac{2\, n_1 n_2 V_{p + 1} \left[(g_1 - g'_1)^2 + (g_2 - g'_2)^2 + (g_1 g'_2 - g_2 g'_1)^2 \right]^{\frac{1}{2}} }{ (8\pi^2 \alpha')^{\frac{1 + p}{2}}} \int_0^\infty \frac{d t}{t^{\frac{9 - p}{2}}}\frac{ e^{ - \frac{y^2}{2\pi \alpha' t}} \, \theta^4_1 \left(\left.\frac{\nu'_0}{2} \right| it \right)}{\eta^9 (it ) \theta_1 (\nu'_0 | it)},\nn
&=& \frac{4\, n_1 n_2 V_{p + 1} \left[(1 + g^2_1 + g^2_2)(1 + g'^2_1 + g'^2_2)\right]^{\frac{1}{2}}  (1 - \cos\pi \nu'_0)^2}{(8\pi^2 \alpha')^{\frac{1 + p}{2}}} \int_0^\infty \frac{d t}{t^{\frac{9 - p}{2}}} e^{ - \frac{ y^2}{2\pi \alpha' t}}\nn
&\,&\times  \prod_{n = 1}^\infty \frac{(1 - 2 |z|^{2n} \cos\pi \nu'_0 + |z|^{4n})^4} {(1 - |z|^{2n})^6 (1 - 2 |z|^{2n} \cos 2\pi \nu'_0 + |z|^{4n})},
\eea 
where in the second equality we have used the $\theta$-function identity  \eqref{theta-identity} and once again $|z| = e^{- \pi t} < 1$. Note that each factor in the integrand of the last equality is positive, therefore this interaction is attractive as expected. One expects to see an open string tachyon mode to appear and for this, we need to pass the above tree-level closed string cylinder to the open string one-loop annulus amplitude via the Jacobi transformation $t \to t' = 1/t$. Using the identities for the Dedekind $\eta$-function and $\theta_1$-function given in \eqref{jacobi}, we have the open string annulus amplitude, from the second equality in \eqref{gg3-totalamplit}, as
\bea\label{gg3-annulusamplit}
\Gamma &=&  - i \frac{2\, n_1 n_2 V_{p + 1} \left[(g_1 - g'_1)^2 + (g_2 - g'_2)^2 + (g_1 g'_2 - g_2 g'_1)^2 \right]^{\frac{1}{2}} }{ (8\pi^2 \alpha')^{\frac{1 + p}{2}}} \int_0^\infty \frac{d t'}{t'^{\frac{ p + 1}{2}}} \,e^{ - \frac{y^2 t'}{2\pi \alpha' }}\nn
&\,&\times  \frac{ \theta^4_1 \left(- \left.\frac{i \nu'_0 t'}{2} \right| it' \right)}{\eta^9 (it' )\, \theta_1 (- i \nu'_0 t' | it')},\nn
&=& \frac{2^4 \, n_1 n_2 V_{p + 1} \left[(g_1 - g'_1)^2 + (g_2 - g'_2)^2 + (g_1 g'_2 - g_2 g'_1)^2 \right]^{\frac{1}{2}} }{ (8\pi^2 \alpha')^{\frac{1 + p}{2}}} \int_0^\infty \frac{d t}{t^{\frac{ p + 1}{2}}}\,  e^{ - \frac{y^2  t}{2\pi \alpha' }} \, \frac{\sinh^4 \frac{\pi \nu'_0 t}{2}}{\sinh\pi \nu'_0 t}\nn
&\,& \times \prod_{n = 1}^\infty \frac{(1 - 2 |z|^{2n} \cosh \pi \nu'_0 t  + |z|^{4n})^4}{(1 - |z|^{2n})^6 (1 - 2 |z|^{2n} \cosh 2 \pi \nu'_0 t + |z|^{4n})},
\eea
where in the second equality we have dropped the prime on $t$ and once again $|z| = e^{- \pi t} < 1$. From the second equality above, we see the factor in the integrand
\be\label{gg3-tachyon}
\lim_{t \to \infty} e^{ - \frac{y^2  t}{2\pi \alpha' }} \, \frac{\sinh^4 \frac{\pi \nu'_0 t}{2}}{\sinh\pi \nu'_0 t} \sim \lim_{t \to \infty} e^{- \frac{(y^2 - 2\pi^2 \nu'_0 \alpha') t}{2\pi \alpha'}}
\ee
which blows up if $y < \pi \sqrt{2 \nu'_0 \alpha'}$, signaling the onset of tachyonic instability. Once again we see that the magnetic fluxes give rise to the open string tachyon mode.

     In summary, we have further confirmed what has been learned in the previous section using systems with different structures of fluxes as discussed in this section. In other words,  the electric flux(es) give rise to the non-perturbative open string pair production while the magnetic one(s) give rise to the tachyon mode. When both of these are present, the interplay of these two gives to the enhancement of the pair production rate.  The pair production enhancement revealed in this section for small $\nu_0$ parameter becomes more suitable in realistic application since it does not necessarily need large fluxes. For this reason, it is more useful as we will discuss in section 6.  In the following, we will discuss the only remaining case for $8 \ge p \ge 4$ which involves only two magnetic fluxes sharing no common field strength index.  As expected, this gives only attractive interaction and a tachyonic instability at small brane separation if we use the open string annulus diagram. 
     
\section{The $ 8 \ge p \ge 4$ case}
  We have only one case to discuss in this section for which we have two magnetic fluxes sharing no common field strength index. The structure of the flux $\hat F$ on one set of Dp branes can be cast without loss of generality as
\be\label{gg4-structure}
\hat F =\left( \begin{array}{ccccccc}
0 & 0 & 0 &0 & 0&0&\ldots\\
0 &0&- g_1&0&0&0&\ldots\\
0&g_1 &0&0&0&0&\ldots\\
0&0&0&0&- g_2&0&\ldots\\
0&0&0&g_2&0&0&\ldots\\
0&0&0&0&0&0&\ldots\\
 \vdots&\vdots&\vdots&\vdots&\vdots&\vdots&\ddots
 \end{array}\right)_{(1 + p)\times (1 + p)},
 \ee
 while on the other set we have the same structure for the flux $\hat F'$ but with a prime to distinguish from the former.  With them, again following the same steps as before, we can determine the corresponding eigenvalues $\lambda_\alpha$ with $\alpha = 0, 1, \cdots, p$ as 
 \bea\label{gg4-eigen1}
 \lambda_0 &=& \lambda_5 = \cdots = \lambda_p = 1, \nn
 \lambda_1 &=& \lambda,  \quad \lambda_2 = \lambda^{-1}, \quad \lambda_3 = \lambda', \quad \lambda_4 = \lambda'^{-1}, 
 \eea
 where $\lambda$ and $\lambda'$ satisfy, respectively, 
 \bea\label{lambda-lambda'}
  \lambda + \lambda^{-1} &=& 2 \frac{(1 - g^2_1)(1 - g'^2_1) + 4 g_1 g'_1}{(1 + g^2_1)(1 + g'^2_1)},\nn
 \lambda' + \lambda'^{-1}  &=&2 \frac{(1 - g^2_2)(1 - g'^2_2) + 4 g_2 g'_2}{(1 + g^2_2)(1 + g'^2_2)}.
 \eea
Since both $\lambda$ and $\lambda'$ are magnetic nature, for the purpose of expressing the interaction amplitude in terms of various $\theta$-functions and the Dedekind $\eta$-functions as before, we set $\lambda = e^{2\pi i \nu'_{10}}$ and $\lambda' = e^{2\pi i \nu'_{20}}$.   Then from \eqref{lambda-lambda'}, we have
\be\label{gg4-nu1nu2}
\tan\pi\nu'_{10} = \frac{|g_1 - g'_1|}{1 + g_1 g'_1},\quad \tan\pi \nu'_{20} = \frac{|g_2 - g'_2|}{1 + g_2 g'_2},
\ee
where $0 < |g_1|,  |g_2|, |g'_1|, |g'_2| < \infty$ but with $0 < \nu'_{10} < 1$ and $0 < \nu'_{20} < 1$, respectively.  In particular, $1/2 > \nu'_{a0} > 0$ for $ - 1 < g_a g'_a < \infty$, 
$\nu'_{a0} = 1/2$ for $g_a g'_a =  - 1$ and $1 > \nu'_{a0} > 1/ 2$ for $ - \infty < g_a g'_a < - 1$. Here $a = 1, 2$, respectively.   The zero-mode matrix element \eqref{0mme} for the present case in the R-R sector can be determined to be
\be\label{gg4-mme}
{}_{\rm 0R}\langle B', \eta'| B, \eta\rangle_{\rm 0R}  = - \frac{ 2^4 \,(1 + g_1 g'_1 )(1 + g_2 g'_2)}{\sqrt{(1 + g^2_1)(1 + g'^2_1)(1 + g^2_2)(1 + g'^2_2) }} \delta_{\eta' \eta, +}.
\ee 
With the above preparation, we can obtain the amplitude in the NS-NS sector as
\bea\label{gg4-amplitudensns}
\Gamma_{\rm NSNS} &=& \frac{1}{2} \left[\Gamma_{\rm NSNS} ( + ) - \Gamma_{\rm NSNS} ( - )\right],\nn
                                    &=&  \frac{n_1 n_2 V_{p + 1} \left[(1 + g^2_1)(1 + g^2_2) (1 + g'^2_1) (1 + g'^2_2) \right]^{\frac{1}{2}}}{2 (8\pi^2 \alpha')^{\frac{1 + p}{2}}} \int_0^\infty \frac{d t}{t^{\frac{9 - p}{2}}} \frac{e^{ - \frac{y^2}{2\pi \alpha' t}}}{ |z|} \nn
    &\,&\times  \left[\prod_{n = 1}^\infty {\cal A}_n ( + ) - \prod_{n = 1}^\infty {\cal A}_n ( - )\right],\nn
 &=&    \frac{2\, n_1 n_2 V_{p + 1} |g_1 - g'_1| |g_2 - g'_2|}{ (8\pi^2 \alpha')^{\frac{1 + p}{2}}} \int_0^\infty \frac{d t}{t^{\frac{9 - p}{2}}} e^{ - \frac{y^2}{2\pi \alpha' t}} \nn
 &\,& \times \frac{\theta^2_3 (0 |it) \theta_3 (\nu'_{10} | it) \theta_3 (\nu'_{20} | it) - \theta^2_4 (0 | it) \theta_4 (\nu'_{10}|it) \theta_4 (\nu'_{20} | it)}{\eta^6 (it) \theta_1 (\nu'_{01} |it) \theta_2 (\nu'_{02} | it)},
 \eea
 where in the first equality we have used \eqref{amplitudensns} for $\Gamma_{\rm NSNS} (\pm)$  and in the second equality ${\cal A}_n (\pm)$ are defined in \eqref{eg3-A} but here with $\lambda$ and $\lambda'$ given  in \eqref{lambda-lambda'}.  Similarly, the amplitude in the R-R sector can be obtained as
 \bea\label{gg4-amplituderr}
 \Gamma_{\rm RR} &=& \frac{1}{2} \left[\Gamma_{\rm RR} ( + ) + \Gamma_{\rm RR} ( - )\right],\nn
 &=& - \frac{2^3\,  n_1 n_2 V_{p + 1} (1 + g_1 g'_1)(1 + g_2 g'_2) }{ (8\pi^2 \alpha')^{\frac{1 + p}{2}}} \int_0^\infty \frac{d t}{t^{\frac{9 - p}{2}}} e^{ - \frac{y^2}{2\pi \alpha' t}}  \prod_{n = 1}^\infty {\cal B}_n,\nn
 &=&  -   \frac{2\, n_1 n_2 V_{p + 1} |g_1 - g'_1| |g_2 - g'_2|}{ (8\pi^2 \alpha')^{\frac{1 + p}{2}}} \int_0^\infty \frac{d t}{t^{\frac{9 - p}{2}}} \frac{e^{ - \frac{y^2}{2\pi \alpha' t}}  \,\theta^2_2 (0 |it) \theta_2 (\nu'_{10} | it) \theta_2 (\nu'_{20} | it)}{\eta^6 (it) \theta_1 (\nu'_{10} | it) \theta_1 (\nu'_{20} | it) },\qquad
 \eea 
 where in the first equality we have used \eqref{amplituderr} for $\Gamma_{\rm RR} (\pm)$ and also \eqref{gg4-mme} for the zero-mode matrix element, and in the second equality ${\cal B}_n$ is defined in \eqref{eg3-B} but again with the present $\lambda$ and $\lambda'$ given in \eqref{lambda-lambda'}.    So the total amplitude
 \bea\label{gg4-totalamplitude}
 \Gamma &=& \Gamma_{\rm NSNS} + \Gamma_{\rm RR}, \nn
&=& \frac{2\, n_1 n_2 V_{p + 1} |g_1 - g'_1| |g_2 - g'_2|}{ (8\pi^2 \alpha')^{\frac{1 + p}{2}}} \int_0^\infty \frac{d t}{t^{\frac{9 - p}{2}}} e^{ - \frac{y^2}{2\pi \alpha' t}}\left[\theta^2_3 (0 |it) \theta_3 (\nu'_{10} | it) \theta_3 (\nu'_{20} | it) - \right. \nn
&& \left. \theta^2_4 (0 | it) \theta_4 (\nu'_{10}|it) \theta_4 (\nu'_{20} | it) - \theta^2_2 (0 |it) \theta_2 (\nu'_{10} |it) \theta_2 (\nu'_{02} | it)\right] /[\eta^{6} (it) \theta_1 (\nu'_{10} | it) \theta_1 (\nu'_{20} | it)],\nn
&=&  \frac{4\, n_1 n_2 V_{p + 1} |g_1 - g'_1| |g_2 - g'_2|}{ (8\pi^2 \alpha')^{\frac{1 + p}{2}}} \int_0^\infty \frac{d t}{t^{\frac{9 - p}{2}}} e^{ - \frac{y^2}{2\pi \alpha' t}}  \frac{\theta^2_1 \left(\left.\frac{\nu'_{10} - \nu'_{20}}{2}\right| it\right) \theta^2_1  \left(\left.\frac{\nu'_{10} + \nu'_{20}}{2}\right| it\right)}{\eta^6 (it)  \theta_1 (\nu'_{10} | it) \theta_1 (\nu'_{20} | it) },\nn
&=&  \frac{4\, n_1 n_2 V_{p + 1} \left[(1 + g^2_1)(1 + g^2_2) (1 + g'^2_1) (1 + g'^2_2) \right]^{\frac{1}{2}}  }{ (8\pi^2 \alpha')^{\frac{1 + p}{2}} (\cos\pi\nu'_{10} - \cos\pi\nu'_{20})^{-2}} \int_0^\infty \frac{d t}{t^{\frac{9 - p}{2}}} e^{ - \frac{y^2}{2\pi \alpha' t}}\nn
&& \times \prod_{n = 1}^\infty \frac{\left[1 - 2 |z|^{2n} \cos\pi (\nu'_{01} - \nu'_{20}) + |z|^{4n}\right]^2 \left[1 - 2 |z|^{2n} \cos\pi (\nu'_{01} + \nu'_{20}) + |z|^{4n}\right]^2}{(1 - |z|^{2n})^4 (1 - 2 |z|^{2n} \cos\pi \nu'_{10} + |z|^{4n}) (1 - 2 |z|^{2n} \cos\pi \nu'_{20} + |z|^{4n})},
\eea
where in obtaining the third equality we have used the identity \eqref{eg3-jacobi} for $\theta$-functions and once again $|z| = e^{- \pi t} < 1$. Note that every factor in the integrand in the last equality is non-negative, so $\Gamma \ge 0$, which vanishes only if $\nu'_{10} = \nu'_{20}$ and otherwise gives an attractive interaction as expected.\\

The small brane separation behavior of the amplitude can be best seen in terms of the open string one-loop annulus amplitude which can be obtained from the third equality of \eqref{gg4-totalamplitude} via the Jacobi transformation $t \to t' = 1/t$. Using the relations for the Dedekind $\eta$-function and the $\theta_1$-function in \eqref{jacobi}, we have the annulus amplitude as
\bea\label{gg4-annulusamplit}
\Gamma &=& - \frac{4\, n_1 n_2 V_{p + 1} |g_1 - g'_1| |g_2 - g'_2|}{ (8\pi^2 \alpha')^{\frac{1 + p}{2}}} \int_0^\infty \frac{d t'}{t'^{\frac{p - 1}{2}}} e^{ - \frac{y^2 t'}{2\pi \alpha' }}  \frac{\theta^2_1 \left(\left.\frac{\nu'_{10} - \nu'_{20}}{2 i} t'\right| it\right) \theta^2_1  \left(\left. \frac{\nu'_{10} + \nu'_{20}}{2 i} t'\right| it\right)}{\eta^6 (it')  \theta_1 (- i\nu'_{10} t' | it') \theta_1 ( - i \nu'_{20} t' | i t') },\nn
&=& \frac{2^4 \, n_1 n_2 V_{p + 1}|g_1 - g'_1| |g_2 - g'_2|}{ (8\pi^2 \alpha')^{\frac{1 + p}{2}}} \int_0^\infty \frac{d t}{t^{\frac{p - 1}{2}}} e^{ - \frac{y^2 t}{2\pi \alpha' }} \frac{\sinh^2\pi\frac{\nu'_{10} - \nu'_{20}}{2} t \sinh^2 \pi \frac{\nu'_{10} + \nu'_{20}}{2} t }{\sinh\pi \nu'_{10} t \sinh\pi \nu'_{20} t}\nn
&\times& \prod_{n = 1}^\infty \frac{\left[1 - 2 |z|^{2n} \cos\pi (\nu'_{10} - \nu'_{20}) t  + |z|^{4n}\right]^2 \left[1 - 2 |z|^{2n} \cos\pi(\nu'_{10} + \nu'_{20}) t + |z|^{4n}\right]^2}{(1 - |z|^{2n})^4 \left[1 - 2 |z|^{2n} \cos2\pi \nu'_{10} t + |z|^{4n}\right] \left[1 - 2 |z|^{2n} \cos 2\pi \nu'_{20} t + |z|^{4n}\right]},\qquad
\eea
where in the second equality we have dropped the prime on $t$ and again $|z| = e^{- \pi t} < 1$.
From this amplitude, it is also clear that $\Gamma  = 0$ only if $\nu'_{10} = \nu'_{20}$ and otherwise it is greater than zero, therefore giving an attractive interaction.  For large $t$, we have an exponentially growing factor if $\nu'_{10} \neq \nu'_{20}$ in the integrand 
\be
\lim_{t \to \infty}  \frac{\sinh^2\pi\frac{\nu'_{10} - \nu'_{20}}{2} t \sinh^2 \pi \frac{\nu'_{10} + \nu'_{20}}{2} t }{\sinh\pi \nu'_{10} t \sinh\pi \nu'_{20} t} \sim e^{\pi |\nu'_{10} - \nu'_{20}| t} \to \infty,
\ee 
which indicates the existence of an open string tachyon mode as expected.  There is a tachyonic instability to occur when $y < \pi \sqrt{2 |\nu'_{10} - \nu'_{20}|\alpha'}$. 

Given what we have learned in the previous sections, the nature of the interaction as well as the onset of tachyonic instability is expected.

\section{Conclusion and discussion}
In this paper, we consider a system of two sets of Dp branes placed parallel at separation with each carrying two worldvolume fluxes. We focus here on that the two fluxes on one set of Dp branes are the same in structure but different in values as those on the other set. We give a systematic account of computing the stringy amplitude for each allowed such system and analyzing the analytical behavior of this amplitude.  We have learned that when the fluxes are electric in nature, they in general give rise to the non-perturbative  Schwinger-type open string pair production. On the other hand, when the fluxes are magnetic in nature, they give rise to an open string tachyon mode and there will be the onset of tachyonic instability and its subsequent tachyon condensation when the brane separation is smaller than a certain value determined by the fluxes.   The interplay of the non-perturbative open string pair production and the tachyon mode leads to the open string pair production enhancement in certain cases when one flux is electric and the other magnetic.  
In particular, we find this enhancement even when the electric flux and the magnetic one share one common field strength as reported in the subcase 3 in subsection 3.2, which is quite unexpected since there is no such enhancement in the one-flux case studied previously by the present author and his collaborator in \cite{Lu:2009au} . This pair production
enhancement can have potential realistic applications which we will discuss later in this section.  

When the two fluxes share one common field strength index, one can examine all the corresponding closed string tree-level cylinder amplitudes computed in the previous sections and find that they can be cast in general as
\be\label{commonindex-cylindera}
\Gamma =   \frac{2\, n_1 n_2 V_{p + 1} [\det(\eta + \hat F')\det(\eta + \hat F)]^{\frac{1}{2}} \sin \pi \nu}{(8 \pi^2\alpha')^{\frac{1 + p}{2}}} \int_0^\infty \frac{d t}{t^{\frac{9 - p}{2}}} \frac{e^{- 
\frac{y^2}{2\pi \alpha' t} }\, \theta^4_1 \left(\left.\frac{\nu}{2}\right| it \right)}{\eta^9 (it) \theta_1 (\nu |it)},
\ee
where the $\nu$ parameter is determined in the previous sections, i.e. \eqref{nuparameter},   \eqref{eg-nuparameter} and \eqref{gg3-nuparameter}, respectively.  When $\nu$ is real, it can be set $\nu = \nu'_0$ with $0 < \nu'_0 < 1$.  The corresponding fluxes are magnetic in nature. The interaction amplitude gives an attractive interaction between the two sets of Dp branes until $y = \pi \sqrt{2\nu'_0 \alpha'}$ when the corresponding open string tachyon condensation occurs. So when the amplitude is expressed in terms of the open string annulus one via the Jacobi transformation $t \to t' = 1/t$, we can see the onset of tachyonic instability by noticing an exponential divergent factor in the integrand of the amplitude by setting $t' \to \infty$ when $y < \pi \sqrt{2\nu'_0 \alpha'}$.  For this case, the only way the system can give off its excess energy due to the applied fluxes is via the tachyon condensation and when this is done, the system becomes 1/2 BPS  just like each set of the system. The necessary condition for this is to have $\Gamma = 0$, which determines the allowed fluxes. 

When $\nu$ is purely imaginary, it can be set $\nu = i \nu_0$ with $0 < \nu_0 < \infty$. The corresponding fluxes are electric in nature. The large brane separation interaction is still attractive but the small brane separation one is rich in physics. In analog of the Schwinger pair production in QED, we know that there will be open string pair production for this case beforehand. This manifests itself again when we express the interaction amplitude in terms of the open string annulus one, implying an imaginary part of the amplitude.  When the brane separation is large, the mass of the open string connecting the two sets of Dp brane, which equals to the string tension times the brane separation, is large, therefore the open string pairs are difficult to be produced from the vacuum.  So for large brane separation, the energy loss due to the pair production can be ignored and the amplitude has almost  no imaginary part.  However, when the brane separation is small, the pair production becomes important and the imaginary part of the amplitude, giving the pair production rate,  can no longer be ignored which can be computed following \cite{Bachas:1992bh}  as we did in the previous sections. The larger the $\nu_0$ is, the larger the pair production rate. In particular, the rate diverges when $\nu_0 \to \infty$, corresponding to the critical electric flux(es).  If the parameter $\nu_0$ is not large, for example, $\nu_0 < 1$,  the pair production rate is in general small even at brane separation $y = 0$ and we may treat the pair production as an adiabatic process until the system becomes again 1/2 BPS one for which the pair production stops.  This can be determined by $\Gamma = 0$ which gives a condition for which the fluxes need to satisfy. Note that the pair production is the process to give off the excess energy of the system before it becomes 1/2 BPS. For this case, there is no open string tachyon mode which appears a bit unexpected since the system itself is not supersymmetric before it becomes 1/2 BPS. One possible explanation to this puzzle is that, unlike the previous magnetic case for which the tachyon condensation serves as an only means to give off the system excess energy  at\footnote{If we extrapolate this to $\nu'_0 = 0$, it would imply that the tachyon condensation occurs at $y < 0$ which is impossible and this may also serve to explain the absence of tachyon mode in the pure electric case.}  $y < \pi \sqrt{2 \nu'_0 \alpha'}$, the pair production gives off the system excess energy at any brane separation to relax the system back to 1/2 BPS one and for this the tachyon mode can hardly manifest itself in the annulus amplitude.   

When $\nu$ is complex as discussed in subcase 3 in subsection 3.2,  we have $\nu = 1 - i \nu_0$ with $0 < \nu_0 < \infty$. This case is impossible when each set of branes carries only one-flux as addressed previously in \cite{Lu:2009au} by the present author and his collaborator.  As discussed in subcase 3 in subsection 3.2, the real part `1' of $\nu$ is actually due to the magnetic fluxes applied while the imaginary part $\nu_0$ is due to both the electric and magnetic fluxes.  The large brane interaction is still attractive while the small brane separation behavior of the amplitude can be best seen as usual in terms of the open string annulus amplitude.  One expects the amplitude  
to have an imaginary part, resulting from an infinite number of simple poles of its integrand, to give rise to  the open string pair production. Moreover the real part unity of $\nu$ gives an enhancement of the pair production and this is the first pair production enhancement reported in this paper which is similar in spirit to the enhancement of pair production discussed in subsection 4.1.  For large $\nu_0$, this enhancement plays less important role and the behavior of the pair production rate is more or less the same as the pure electric case discussed above.   The most interesting and useful case is for small but fixed $\nu_0$ for which the enhancement is important.  The pair production rate \eqref{eg-nu02rate} can now be approximated as 
\be\label{cdeg-pprate} 
(2 \pi \alpha')^{\frac{1 + p}{2}} {\cal W} \approx \frac{2 \pi n_1 n_2 \sqrt{(1 + a^2) (1 + a'^2)} }{(4 \pi)^{\frac{1 + p}{2}}} \sum_{k = 1}^\infty (- )^{k - 1} \left(\frac{\nu_0}{k}\right)^{\frac{1 + p}{2}}e^{- \frac{k (y^2 - 2\pi^2 \alpha')}{2\pi \nu_0 \alpha'}},
\ee
where we have set $g = a \cosh\theta, f = a \sinh\theta, g' = a' \cosh\theta', f' = a' \sinh\theta'$.  For small $\nu_0$, this rate can be possibly significant if $y = \pi \sqrt{2 \alpha'} + 0^+$ and both $|a|$ and $|a|'$ are large. Further the smallest allowed $p = 2$ gives the largest rate when the fluxes are taken the same for all $2 \le p \le 8$. This can be interesting academically but in potentially realistic applications, we cannot have large $|g|$ or $|g'|$ or both since large $|a|$ or $|a'|$ or both imply them. Further if $\nu_0 \ll 1$, the above rate cannot be significant even with large $|a|$ and $|a'|$.  However, there is an exception if we are allowed to have the brane separation $y < \pi \sqrt{2\alpha'}$. If so, large magnetic fluxes are not needed to have a significant pair production rate. The rational for the present case is the same as for the case  when the electric flux and the magnetic one do not share a common field strength spatial index which we will turn next.  So we will leave this discussion in appropriate place there.      

When the two fluxes share no common field strength index, the closed string cylinder amplitude can also be cast in general as 
\bea\label{nocommonindex-cylindera}
\Gamma &=&   \frac{2^2\, i  n_1 n_2 V_{p + 1} [\det(\eta + \hat F')\det(\eta + \hat F)]^{\frac{1}{2}} \sin \pi \nu \sin \pi \nu'}{(8 \pi^2\alpha')^{\frac{1 + p}{2}}} \int_0^\infty \frac{d t}{t^{\frac{9 - p}{2}}} e^{- 
\frac{y^2}{2\pi \alpha' t} }\nn
&\,& \times \frac{ \theta^2_1 \left(\left.\frac{\nu - \nu'}{2}\right| it \right)\theta^2_1 \left(\left.\frac{\nu + \nu'}{2}\right| it \right)}{\eta^6 (it) \theta_1 (\nu |it) \theta_1 (\nu' | it)},
\eea
where the $\nu$ and $\nu'$ parameters are determined in the previous sections, i.e. \eqref{eg3-nu-nu'} and  \eqref{gg4-nu1nu2}, respectively.    
 
 For large brane separation, all the systems considered have a well-defined finite attractive interaction for $p < 7$. When both $\nu$ and $\nu'$ are real with $0 < \nu < 1$ and $0 < \nu' < 1$, the corresponding fluxes are all magnetic. For this case, the interaction amplitude is positive, implying attractive interaction, until the brane separation $ y = \pi \sqrt{2 |\nu - \nu'| \alpha'}$ for which the open string tachyon condensation occurs. As before, the onset of tachyonic instability can be best seen in terms of the open string annulus amplitude and 
 the tachyon condensation once again serves as the only means to give off the excess energy of the system to finally settle it down to its 1/2 BPS state. 
 
 When one flux is electric and the other is magnetic, we have, say, $\nu = i \nu_0$ with $0 < \nu_0 < \infty$ and $\nu' = \nu'_0$ with $0 < \nu'_0 < 1$. Once again we expect a significant open string pair production at small brane separation and the pair production rate is given in general by \eqref{eg3-pprate}. The large $\nu_0$ case is not different from the previous cases and once again the magnetic flux plays a minor role. We focus here on the realistic useful case for which we have small but fixed $\nu'_0$ and $\nu_0$ with $\nu'_0/\nu_0 \gg 1$. For this, the dimensionless pair production rate can be approximated from \eqref{eg3-pprate} as 
\be\label{cdeg3-approx-pprate}
(2\pi \alpha')^{\frac{1 + p}{2}} {\cal W}  (\nu'_0 \neq 0) = \frac{n_1 n_2  |f - f'||g - g'|}{2\pi^2} \sum_{k = 1}^\infty (-)^{k - 1} \frac{1}{k} \left(\frac{\nu_0}{4 \pi k}\right)^{\frac{p - 3}{2}}  e^{- \frac{ k\, y^2}{2\pi \alpha' \nu_0}}\, e^{\frac{k \pi  \nu'_0}{\nu_0}}.
\ee    
 Small $\nu_0$ implies small $|f - f'|$. In other words, the electric flux on one set of Dp branes is almost identical to that on the other set of Dp branes. Since $0 < \nu'_0 < 1$, so small but fixed $\nu'_0$ does not necessarily imply small magnetic fluxes $|g|$ and $|g'|$. As our sample estimation demonstrates in Case III in subsection 4.1, the largest rate is for $p = 3$ and can be significant for a reasonable choice of $\nu_0, \nu'_0$ when $y = \pi \sqrt{2 \nu'_0 \alpha'} + 0^+$.  This appears that we can have a real experimental possibility for exploring the existence of extra dimension(s) and as such for testing string theories if we assume to live in a (1 + 3)-dimensional world which are D3 branes.  As discussed and stressed in Introduction and at various points in the previous sections, the open string pair production gives rise to the open string pairs connecting the two sets of Dp branes and therefore they are directly related to the existence of extra dimension(s). Since this is based on string theories,  a detection of open string pair production indicates not only the existence of extra dimension(s) but also the correctness of string theories.   For an observer living on one set of D3 branes, she/he can only detect the ends of each produced open string pair as a particle/anti-particle pair. However, there is a sharp distinction between the pair production here and the Schwinger pair production, say, in QED.  In string theories, if the set of D3 branes carrying the same fluxes is an isolated one, the observer living on this set will not detect pair production since a charged neutral string with its both ends on the D3 branes will not give rise to the pair production.   While this is not the case for Schwinger pair production, say, in QED.  Further the Schwinger pair production on the magnetic flux, 
 even the non-linear effect is considered,  is different. 
 
 If we indeed want to put the above into test in real-life experiment, for example, in real-life laboratory, the electric flux and the magnetic flux are both very small compared to the string scale even with the consideration of some D-brane phenomenological one, say, around 10 TeV.  In what follows, we still want to keep $\nu'_0 /\nu_0 \gg 1$. One of the good things for such small fluxes is that we do not expect the pair production along with possible tachyon condensation to perturb the original brane system much. Their roles are to release the tiny excess energy, due to the applied fluxes, in comparing with the rest energy of the original system. So we may expect that the pair production rate can be valid even to zero brane separation.  Let us demonstrate this in the most useful case of $p = 3$ and for this case the rate \eqref{cdeg3-approx-pprate} is
 \be\label{cdeg3p3-pprate}
  (2\pi \alpha')^2 {\cal W}  (\nu'_0 \neq 0) = \frac{n_1 n_2  |f - f'||g - g'|}{2\pi^2} \sum_{k = 1}^\infty (-)^{k - 1} \frac{1}{k}  e^{- \frac{ k\, y^2}{2\pi \alpha' \nu_0}}\, e^{\frac{k \pi  \nu'_0}{\nu_0}}.
\ee    
Even though the large $k$ term in \eqref{cdeg3p3-pprate} appears divergent for $y < \pi \sqrt{2\nu'_0 \alpha'}$,  indicating the tachyonic instability, the rate itself is actually finite if we perform the summation  and the result is 
 \be \label{cdp3sum-pprate}
 (2\pi \alpha')^2 {\cal W}  (\nu'_0 \neq 0) = \frac{n_1 n_2  |f - f'||g - g'|}{2\pi^2}  \ln \left(1 + e^{- \frac{y^2 - 2\pi^2 \nu'_0 \alpha'}{2\pi \nu_0 \alpha'}} \right).
 \ee
 For large brane separation, this rate gives the leading $k = 1$ term approximation of \eqref{cdeg3p3-pprate},  as expected, which remains as a reasonably good approximation until $y = \pi \sqrt{2\nu'_0 \alpha'}$. There is no divergence in this rate even for $y < \pi \sqrt{2\nu'_0 \alpha'}$ down to $y = 0$, giving a fair justification of our above assertion. For $y = 0$, we have the rate as $(2\pi \alpha')^2 {\cal W}  (\nu'_0 \neq 0) = n_1 n_2 |g - g'|^2 /(2\pi)$, which can be significant in terms of laboratory scale but needs further understanding.   As discussed in \cite{Lu:2017tnm}, the rate for $p > 3$ is at least smaller by an order of $(\nu_0 /4\pi)^{1/2}$, which can be a few order of magnitude smaller for $\nu_0$ using real-life laboratory electric flux(es). So the detection of open string pair production can single out D3 branes as the most preferable to its observer, if he/she just like us knows about string theory.  The produced large number of open string pairs can
in turn annihilate to give, for example, highly concentrated high energy photons if the fluxes are localized on the branes and this may have observational consequence such as the Gamma-ray burst.  This same type of pair production and its subsequent annihilation, if happens at our early Universe,  may also be useful in providing a new mechanism for reheating process after cosmic  inflation. 

\section*{Acknowledgements} 
The author acknowledges support by grants from the NSF of China with Grant No: 11235010 and 11775212.


\end{document}